\documentclass[a4paper,prl,twocolumn,showpacs,superscriptaddress,floatfix]{revtex4-2}

\usepackage{graphicx}
\usepackage{dcolumn}
\usepackage{bm}
\usepackage{hyperref}
\hypersetup{
  colorlinks=true,        
  linkcolor=magenta,         
  citecolor=cyan,         
  urlcolor=blue           
}

\usepackage{colortbl}
\usepackage{amsmath}
\usepackage{times}
\usepackage{times}
\usepackage{multirow}
\usepackage[normalem]{ulem}
\usepackage{academicons}
\usepackage{xcolor}
\newcommand{\orcid}[1]{\href{https://orcid.org/#1}{\textcolor[HTML]{A6CE39}{\includegraphics[scale=0.06]{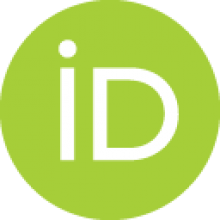}}}}

\definecolor{revcolor}{RGB}{150,50, 70}

\usepackage{acro}

\DeclareAcronym{ns}{
  short=NS,
  long=neutron star,
}

\DeclareAcronym{nsm}{
  short=NSM,
  long=neutron star matter,
}
\DeclareAcronym{pnm}{
  short=PNM,
  long=pure neutron matter,
}

\DeclareAcronym{bns}{
  short=BNS,
  long=binary neutron star,
}

\DeclareAcronym{eos}{
  short=EOS,
  long=equation of state,
}

\DeclareAcronym{gws}{
  short=GWs,
  long=gravitational waves,
}

\DeclareAcronym{qcd}{
  short=QCD,
  long=quantum chromodynamics,
}

\DeclareAcronym{lqcd}{
  short=lQCD,
  long=lattice quantum chromodynamics,
}

\DeclareAcronym{pqcd}{
  short=pQCD,
  long=perturbative quantum chromodynamics,
}

\DeclareAcronym{qnm}{
  short=QNM,
  long=quasi-normal mode,
}

\DeclareAcronym{ur}{
  short=UR,
  long=universal relation,
}

\DeclareAcronym{bps}{
  short=BPS,
  long=Bethe-Pethick-Sutherland,
}

\DeclareAcronym{ddb}{
  short={DDB},
  long={a nucleonic $\beta-$ equilibritated EOS based on a relativistic description of hadrons through their density-dependent couplings  constrained by the existing observational, theoretical and experimental data through a Bayesian analysis},
}

\DeclareAcronym{ddbhyb}{
  short={DDB-Hyb},
  long={a hybrid set of EOSs which consists of the DDB EOS at low density ($\leq 2\rho_0$) and the deconfined quark matter at very high densities ($\geq 40\rho_0$) while the region ($ 2\rho_0$-$ 40\rho_0$) is interpolated by piecewise polytropes},
}

\DeclareAcronym{cft}{
  short=chEFT,
  long=chiral effective field theory,
}

\DeclareAcronym{rmf}{
  short=RMF,
  long={relativistic mean field},
}

\DeclareAcronym{tov}{
  short={TOV},
  long={Tolman-Oppenheimer-Volkoff},
}

\DeclareAcronym{gr}{
  short={GR},
  long={general relativistic},
}

\begin{document}

\title{Robust universal relations in neutron star asteroseismology}

\author{Deepak Kumar \orcid{0000-0001-9292-3598}} 
\affiliation{Theory Division, Physical Research Laboratory, Navarangpura, Ahmedabad 380 009, India}
\affiliation{Indian Institute of Technology Gandhinagar, Gandhinagar 382 355, Gujarat, India }
\email{deepakk@prl.res.in }

\author{Tuhin Malik \orcid{0000-0003-2633-5821}}
\affiliation{CFisUC, Department of Physics, University of Coimbra, PT 3004-516 Coimbra, Portugal}
\email{tuhin.malik@uc.pt}

\author{Hiranmaya Mishra  \orcid{0000-0001-8128-1382}}
\affiliation{Theory Division, Physical Research Laboratory, Navarangpura, Ahmedabad 380 009, India}
\email{hm@prl.res.in}
\affiliation{School of Physical Sciences, National Institute of Science Education
 and Research, Jatni, 752050 India}
\email{hiranmaya@niser.ac.in}

\author{Constan\c ca Provid\^encia  \orcid{0000-0001-6464-8023}}
\affiliation{CFisUC, Department of Physics, University of Coimbra, PT 3004-516 Coimbra, Portugal}
\email{cp@uc.pt}

\date{\today}
 
\begin{abstract}
The non-radial oscillations of the neutron stars (NSs) have been suggested as an useful tool to probe the composition of neutron star matter (NSM). With this scope in mind, we consider a large number of equations of states (EOSs) that are consistent with nuclear matter properties and pure neutron matter EOS based on a chiral effective field theory (chEFT) calculation for the low densities and perturbative QCD EOS at very high densities. This ensemble of EOSs is also consistent with astronomical observations, gravitational waves in GW170817, mass and radius measurements from Neutron star Interior Composition ExploreR (NICER). We analyze the robustness of known universal relations (URs) among the quadrupolar $f$ mode frequencies, masses and radii with such a large number of EOSs and we  find a new UR that results from a  strong correlation between the $f$ mode frequencies and the radii of NSs. Such a correlation is very useful in accurately determining the radius from a measurement of $f$ mode frequencies in the near future. We also show that the quadrupolar $f$ mode frequencies of NS of masses 2.0 M$_\odot$ and above lie in the range $\sim$ 2-3 kHz in this ensemble of physically realistic EOSs. A NS of mass 2M$_{\odot}$ with a low $f$ mode frequency may indicate the existence of non-nucleonic degrees of freedom.
\end{abstract}
\maketitle

\textit{Introduction.} \label{sec:introduction}
The \ac{ns} observations in the multi-messenger astronomy have piqued a lot of interest in the field of nuclear astrophysics and strong interaction physics. The recent radio, x-rays and \ac{gws} observations in the context of \ac{ns}s have provided interesting insights into the properties of matter at high density. The core of such compact objects is believed to contain matter at few times nuclear saturation density, $\rho_0$ ($\rho_0 \approx$ 0.16 fm$^{-3}$) \cite{book.Glendenning1996, book.Haensel2007, Rezzolla:2018jee, Schaffner-Bielich:2020psc} and provides an unique window to get an insight into the behavior of matter at these extreme densities. On the theoretical side, no controlled reliable calculations are there that can be applicable to matter densities relevant for the \ac{ns} cores. The \ac{lqcd} simulations are challenging at these densities due to sign problem in Monte-Carlo simulations. On the other hand, the analytical calculations like \ac{cft} is valid only at low densities while \ac{pqcd} is reliable at extremely high densities. In recent approaches, the \ac{eos}s between these two limits have been explored by connecting these limiting cases using a piecewise polytropic interpolation, speed of sound interpolation or spectral interpolation \cite{Lindblom2012, Kurkela:2014vha, Most:2018hfd, LopeOter:2019pcq, Annala2019, Annala:2021gom, Altiparmak:2022bke}.

The \ac{ns} properties such as mass, radius and quadrupole deformation of merging \ac{ns}s can constrain the uncertainty in \ac{eos}. The discovery of massive \ac{ns} with masses of the order of $2 M_{\odot}$ requires the \ac{eos} to be stiff. However, the fact that non-nucleonic degrees of freedom soften the \ac{eos} at high density,  puts a constraint on the \ac{eos} at the intermediate densities. The observations of \ac{gws} from \ac{bns} inspiral by Advanced LIGO and Advanced Virgo \ac{gws} observatories have opened a new window in the field of multi-messenger astronomy and nuclear physics. The inspiral phase of \ac{ns}-\ac{ns} merger leads to tidal deformation ($\Lambda$), which is strongly sensitive to the compactness. Since $\Lambda$ is related to the \ac{eos} of the \ac{nsm}, this measurement acts as another constraint on the \ac{eos}. On the other hand, recovering the nuclear matter properties from the \ac{eos} of $\beta$-equilibriated matter is rather non trivial. This further requires the knowledge of the composition ({\it e.g.} proton fraction) of matter at high densities \cite{Tovar2021, Imam2021, Mondal2021, Essick:2021vlx}.

In the context of \ac{gws}, the non-radial oscillations of \ac{ns} are particularly interesting as they can carry information of the internal composition of the stellar matter. These oscillations in the presence of perturbations (electromagnetic or gravitational) can emit \ac{gws} at the characteristic frequencies of its \ac{qnm}. The frequencies of \ac{qnm} depend on the internal structure of \ac{ns} and it may be another probe to get an insight regarding the composition of \ac{nsm} also known as asteroseismology. Different \ac{qnm}s are distinguished by the restoring forces that act on the fluid element when it gets displaced from its equilibrium position. The important fluid modes related to \ac{gws} emission include  fundamental $(f)$ modes, pressure $(p)$ modes and gravity $(g)$ modes driven by the pressure and buoyancy respectively. The frequency of $p$ modes is higher than that of $g$ modes while the frequency of $f$ modes lies in between. The focus of the present investigation is on the quadrupolar $f$ modes that are correlated with the tidal deformability during the inspiral phase of \ac{ns} merger \cite{Chirenti:2012wn} and have the strongest tidal coupling among all the oscillation modes. More importantly, these modes lie within the sensitivity range of the current as well as upcoming generation of the \ac{gws} detector networks \cite{Pratten:2019sed}. In this context, \ac{qnm}s have been studied with various \ac{eos} models and some universal/quasi-universal behaviors for the frequency and damping time which are insensitive to the \ac{eos} models \cite{Andersson:1996pn, Andersson:1997rn, Benhar:1998au, Benhar:2004xg, Tsui:2004qd, Chan:2014kua, Sotani:2021nlx, Sotani:2021kiw}. This needs to be explored further regarding the robustness of these relations for a large number of \ac{eos}s consistent with recent observational constraints.

In this letter we propose two major points of interest. Firstly we estimate, within the Cowling approximation \cite{McDermott:1983apj, Yoshida:2002vd}, the $f$ mode oscillation frequencies for \ac{ns}s using a large number of \ac{eos}s and demonstrate that observation of $f$ mode frequencies, apart from causality $c_s^2\leq 1$ and maximum mass constraints, further restrict the \ac{eos}s. Secondly we verify the robustness of few \ac{ur} among the quadrupolar $f$ mode frequencies, masses and radii studied earlier with limited \ac{eos}s. It has been earlier found that these \ac{ur}s between \ac{ns} properties are strongly violated by hybrid \ac{eos}s \cite{Lau:2018mae, Bandyopadhyay:2017dvi, Han:2018mtj} and certain exotic phases \cite{vonDoetinchem:2020vbj}. We consider here a large number of \ac{eos}s and confirm that a known UR is almost insensitive to the \ac{eos}s, while a second one depends slightly on the composition of the \ac{eos}s, i.e. the presence or not of non-nucleonic degree of freedom, and, finally,  we propose a new UR.

{\it Setup - } \label{sec:setup}
The two ensembles of \ac{eos}s that we consider here are constructed by stitching together \ac{eos}s valid for different segments in baryon densities. For the outer crust the \ac{bps} \ac{eos} is chosen \cite{Baym:1971pw}. Outer crust and the core are joined using a polytropic form $p(\varepsilon)=a_1 + a_2 \varepsilon^{\gamma}$ in order to construct the inner crust, where the parameters $a_1$ and $a_2$ are determined in such a way that the \ac{eos} for the inner crust matches with the outer crust at one end ($\rho=10^{-4}$ fm$^{-3}$) and with the core at the other end ($\rho=0.04$ fm$^{-3}$). The polytropic index $\gamma$ is taken to be $4/3$ \cite{Carriere:2002bx}. It is important to note that the differences in \ac{ns}s radii between this treatment of the inner crust \ac{eos} and the unified inner crust description including the pasta phases have been found to be less than 0.5 km, as discussed in \cite{Malik:2022zol}.
The core \ac{eos}s are considered within  two different approaches: 
(i) \ac{ddb}, obtained in \cite{Malik:2022zol}, which satisfies \ac{pnm} constraints at low densities obtained from next-to-next-to-next-to leading order (N$^3$LO) calculations in the \ac{cft} \cite{Tews:2012fj, Hebeler:2013nza}.
(ii) \ac{ddbhyb}.
For the deconfined quark matter, we employ NNLO \ac{pqcd} results of Refs. \cite{Kurkela:2014vha, Kurkela:2009gj} which can be cast in a simple fitting function for the pressure as a function of chemical potential ($\mu$) given as
\begin{align}
P_{pQCD}(\mu) = \frac{\mu^4}{108\pi^2}\left(c_1 - \frac{d_1X^{-\nu_1}}{(\mu/GeV)-d_2X^{-\nu_2}}\right)
\label{eq:pqcd}
\end{align}
where the parameters are $c_1 = 0.9008$, $d_1 = 0.5034$, $d_2 = 1.452$, $\nu_1 = 0.3553$ and $\nu_2 = 0.9101$ \cite{Fraga:2013qra}. Here $X$ is a dimensionless renormalization scale parameter, $X=3\bar{\Lambda}/\mu$ which is allowed to vary $X \in \left[1,4\right]$. We use this \ac{pqcd} \ac{eos} for densities beyond $\rho \simeq 40 \rho_0$ which corresponds to $\mu_{\rm pQCD}=2.6$ GeV \cite{Fraga:2013qra}. Between the region of the validity of \ac{pqcd} and \ac{ddb} i.e. $\mu_{\rm DDB} \leq \mu \leq \mu_{\rm pQCD}$, where $\mu_{\rm DDB}$ is the chemical potential of \ac{ddb} \ac{eos} at $\rho=2\rho_0$, we divide the interval into two segments, ($\mu_{\rm DDB}$-$\mu_c$) and ($\mu_c$-$\mu_{\rm pQCD}$), and assume \ac{eos} has a polytropic form in each segment i.e. $P_i(\rho_i)=\kappa_i \rho_i^{\gamma_i}$ for the {\it i}-th segment \cite{Kurkela:2009gj}. The segments can be connected to each other by requiring that the pressure and  the energy density are continuous at $\mu_c$ as well as the pressure should be an increasing function of the energy density and the \ac{eos} must be subluminal. We also ensure that there is no jump in the baryon number density. This corresponds to assuming no first order phase transition between hadronic matter and quark matter. If one wishes to include a first order phase transition, an extra term to the number density at $\mu_c$ can be added \cite{Kurkela:2009gj}.

To obtain the \ac{eos} of the core, we proceed as follows. For the outer core, which extends approximately until $\rho=2\rho_0$, we use a soft (stiff) \ac{ddb} \ac{eos} as obtained in Ref. \cite{Malik:2022zol} within 90\% CI. The corresponding value of chemical potential at $\rho=2\rho_0$ is $\mu_{\rm DDB}=1.036\ (1.097)$ GeV for a soft (stiff) \ac{ddb} \ac{eos}. We interpolate the region from $\mu=\mu_{\rm DDB}$ to $\mu=\mu_c$ and from $\mu=\mu_c$ to $\mu=\mu_{\rm pQCD}$ with two piecewise polytropes. We select all those \ac{eos}s which (i) match with \ac{pqcd} at $\mu=\mu_{\rm pQCD}$ (i.e. $X\in[1,4]$) (ii) have pressure as an increasing function of energy density, and (iii) are subluminal. We refer this \ac{eos} as \ac{ddbhyb}. The chemical potential $\mu_c$ is here chosen in such a way that the EOS matches \ac{pqcd} at $\mu=\mu_{\rm pQCD}$. We take $\mu_c \in [1.04,2.2]$ GeV and the corresponding pressure $P_c\in[20,1260]$ MeV.fm$^{-3}$. For an ensemble of \ac{ddbhyb} \ac{eos}s we choose $\mu_c,P_c$ randomly in the prescribed domain by Latin-Hypercube-Sampling method \cite{FLORIAN1992123} for an uniform distribution. For a given $\mu_c,P_c$ and $P_{\rm DDB}$, the parameters of the first polytrope, $(\kappa_1,\gamma_1)$ get determined. Similarly for a given $\mu_c,P_c$ and $P_{2}$ (where $P_{2}$ is the \ac{pqcd} pressure for a given value of $X$ at $\mu=\mu_{\rm pQCD}$), the parameters of the second polytrope ($\kappa_2,\gamma_2$) get determined. The domains for pressure ($P_c$) and chemical potential ($\mu_c$) become $P_c \in [45,1255]$ MeV$\cdot$fm$^{-3}$ and $\mu_c \in [1.07,2.09]$ GeV after constrained by \ac{pqcd}. These domains further squeeze to $P_c\in[53,680]$ MeV.fm$^{-3}$ and $\mu_c\in[1.15,1.88]$ GeV after putting the constraint of $M_{\rm max} \geq 2 M_{\odot}$ and so we find 0.38 million \ac{eos}s out of 54 million sampled \ac{eos}s satisfying these constraints. It may be mentioned here that, although we use two polytropes for the interpolation between ($\mu_{\rm DDB}$-$\mu_{\rm pQCD}$), there have been different interpolation functions like spectral decomposition \cite{Lindblom:2010bb, Lindblom:2018rfr} and speed of sound method \cite{Annala2019, Annala:2019puf, Altiparmak:2022bke}.

{\it Pulsating equations -}
To estimate the specific oscillation frequency of \ac{ns}s, let us discuss the non-radial oscillation of a spherically symmetric \ac{ns} characterized by the background space-time metric where the line element is given by 
\begin{align}\label{eq:spherically_symmetric_metric}
ds^2=-e^{2\Phi}dt^2+e^{2\Lambda}dr^2+r^2\left(d\theta^2+\sin^2\theta d\phi^2\right).
\end{align}
We shall consider the pulsating equations within the Cowling approximation so that our study is limited to the modes related to fluid perturbations and neglect the metric perturbations. The Lagrangian fluid displacement vector is given by 
\begin{align}
\xi^i = \left(e^{-\Lambda}W,~ -V\partial_{\theta},~ -V\sin^{-2}\theta \partial_{\phi}\right)r^{-2}Y_{lm}
\end{align}
Where $W(r,t)$ and $V(r,t)$ are the perturbation functions and $Y_{lm}$ are the spherical harmonic function. The perturbation equations that describe oscillations can be obtained by the perturbed Einstein field equations $\delta G_{\alpha\beta} = 8\pi\delta T_{\alpha\beta}$ with $G_{\alpha\beta} = R_{\alpha\beta} - \frac{1}{2}g_{\alpha\beta}R$ being the Einstein tensor. Linearising these equations in the perturbation, while choosing a harmonic time dependence for the perturbation i.e. $W(r,t) \propto W(r)e^{i\omega t}$ and $V(r,t) \propto V(r)e^{i\omega t}$ with frequency $\omega$, the differential equations further fluid perturbation functions can be obtained as \cite{McDermott:1983apj, Sotani:2021nlx, Kumar:2021hzo}
\begin{align} \label{eq:w}
W^{\prime} &= \frac{d\epsilon}{dP}\Big( \omega^2 r^2 e^{\Lambda-2\Phi} V + W \Phi^{\prime} \Big)-l(l+1)e^{\Lambda}V, \\
V^{\prime} &= 2V \Phi^{\prime}-\frac{1}{r^2}W e^{\Lambda},
\end{align}
here, the `prime' denotes the total derivative with respect to $r$. These equations are solved with appropriate boundary conditions at the stellar center $r=0$ and at the surface $r=R$. The $W$ and $V$ in the vicinity of the stellar center are taken as $W(r) \sim Cr^{l+1}$ and $V(r) \sim - Cr^l/l$, where $C$ is an arbitrary constant. The other boundary condition that needs to be full-filled is that the Lagrangian perturbation to the pressure must vanish at the stellar surface. This leads to \cite{Sotani:2021nlx, Kumar:2021hzo, McDermott:1983apj} 
\begin{align}
\omega^2 r^2 e^{\Lambda-2\Phi} V+ W \Phi^{\prime}\big|_{r=R}=0
\end{align}
This apart in the case of density discontinuity these equations have to be supplemented by an extra junction condition at the surface of discontinuity. We shall not consider here a density discontinuity. With these boundary conditions, the problem becomes an eigenvalue problem for the parameter $\omega$ which can be estimated numerically. We shall confine ourselves to $l=2$ quadrupolar modes.

\begin{figure}
    \centering
        \includegraphics[width=0.42\textwidth,angle=0]{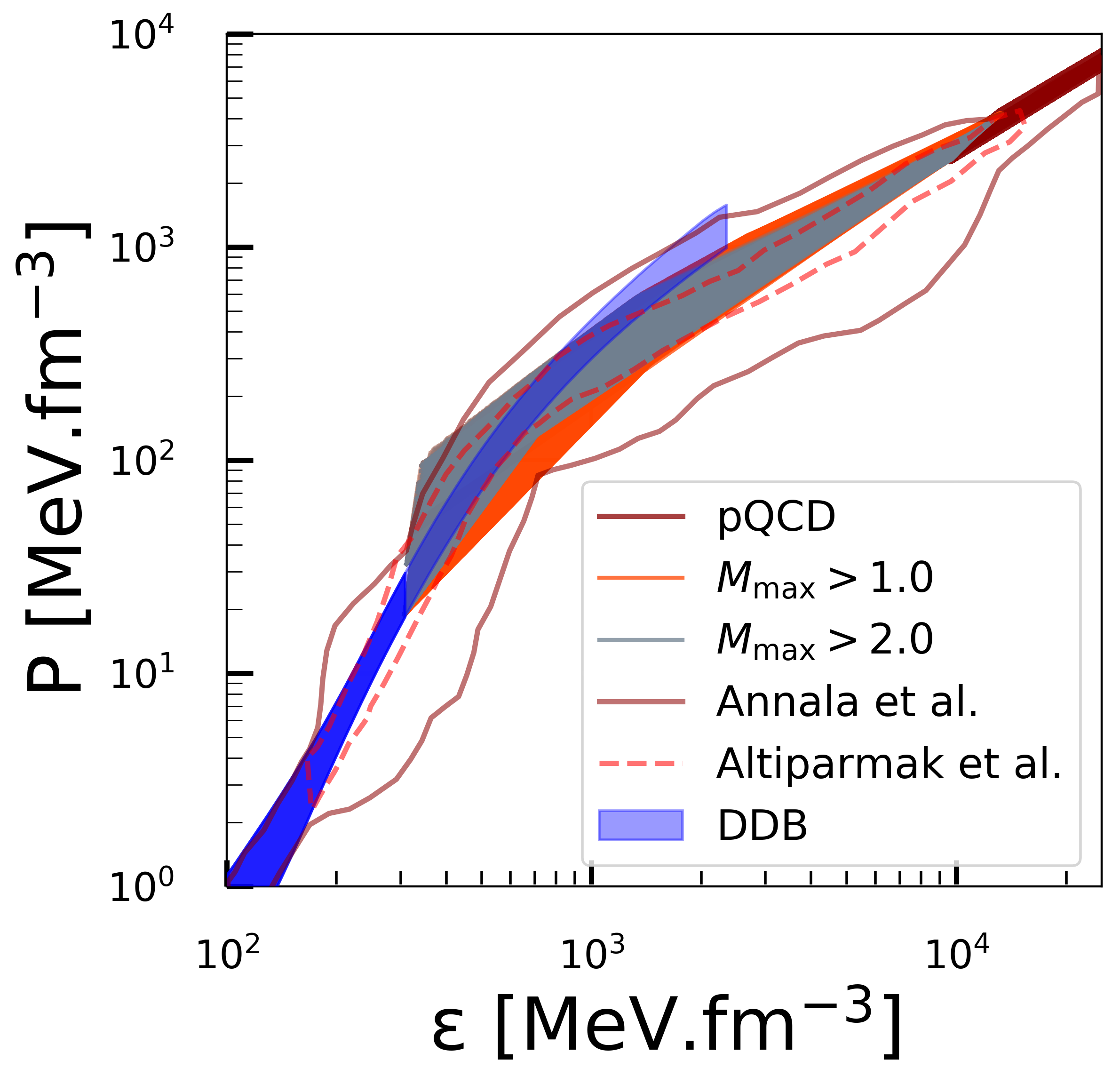}
\caption{We show pressure and energy density regions in MeV.fm$^{-3}$ of our sampled EOSs (\ac{ddb} and \ac{ddbhyb}). We consider nucleonic $\beta$-equilibrated \ac{eos} of the 90\% CIs for \ac{ddb} (light blue) as a full range and (dark blue) up to $2\rho_0$ \cite{Malik:2022zol} and at very high density $\sim 40 \rho_0$ the NNLO pQCD (dark red) \cite{Kurkela:2014vha}. In the intermediate region, \ac{eos} is evolved in thermodynamically consistent way with two polytropic segments (see text for details).  Also included are the limits of the domain of \ac{eos}s obtained in Ref. \cite{Annala:2021gom} (red solid curve) and  the dense PDF ($\geq 0.08$) calculated in Ref. \cite{Altiparmak:2022bke} (red dashed lines). \label{fig:eos_and_g1g2}}
\end{figure}

{\it Results -} \label{sec:results}
We now proceed to analyze the ensembles of \ac{eos}s that are consistent with nuclear matter properties or \ac{pnm} \ac{eos} based on theoretically robust \ac{cft} at low densities and \ac{pqcd} at very high densities. As mentioned earlier, we start with $54$ million \ac{eos}s. We discard those \ac{eos} which do not match the two end points or are superluminal (square of speed of sound $c_s^2 > 1$) as well as the condition of positive speed of sound. This leaves us with an ensemble of 0.38 million \ac{ddbhyb} \ac{eos}s. This ensemble of \ac{eos}s is represented in Fig. \ref{fig:eos_and_g1g2} by the orange band. We next enforce the $M_{\rm max} \geq 2.0M_{\odot}$ constraint resulting from solving the \ac{tov} equations with this ensemble. This constraint further reduces the number of \ac{eos}s to 55,000 which are displayed in Fig. \ref{fig:eos_and_g1g2} as the gray band, named here after \ac{ddbhyb} set. The polytrope indices $\gamma_1$ and $\gamma_2$ are seen to vary over an intervals $\gamma_1 \in [1.67,13.76]$ and $\gamma_2 \in [1.0,1.51]$. The tight constraint on $\gamma_2$ has its origin on the matching to the \ac{pqcd} pressure. In Fig. \ref{fig:eos_and_g1g2}, the light blue band is the $\beta$-equilibrated nuclear matter $\approx 10$k \ac{eos}s (\ac{ddb} 90\% CI) while the dark red band corresponds to \ac{pqcd} \ac{eos}. For comparison, we also plot the domain of \ac{eos}s obtained in Ref. \cite{Annala:2021gom} (red solid curve) compatible with recent NICER and \ac{gws} observations. The red dashed lines refers to the dense PDF ($\geq 0.08$) obtained in Ref. \cite{Altiparmak:2022bke} with continuous sound speed and consistent not only with nuclear theory and \ac{pqcd}, but also with astronomical observations. It is to be noted that both of \ac{ddb} and \ac{ddbhyb} sets are compatible with them.
\begin{figure}
    \includegraphics[width=0.48\textwidth]{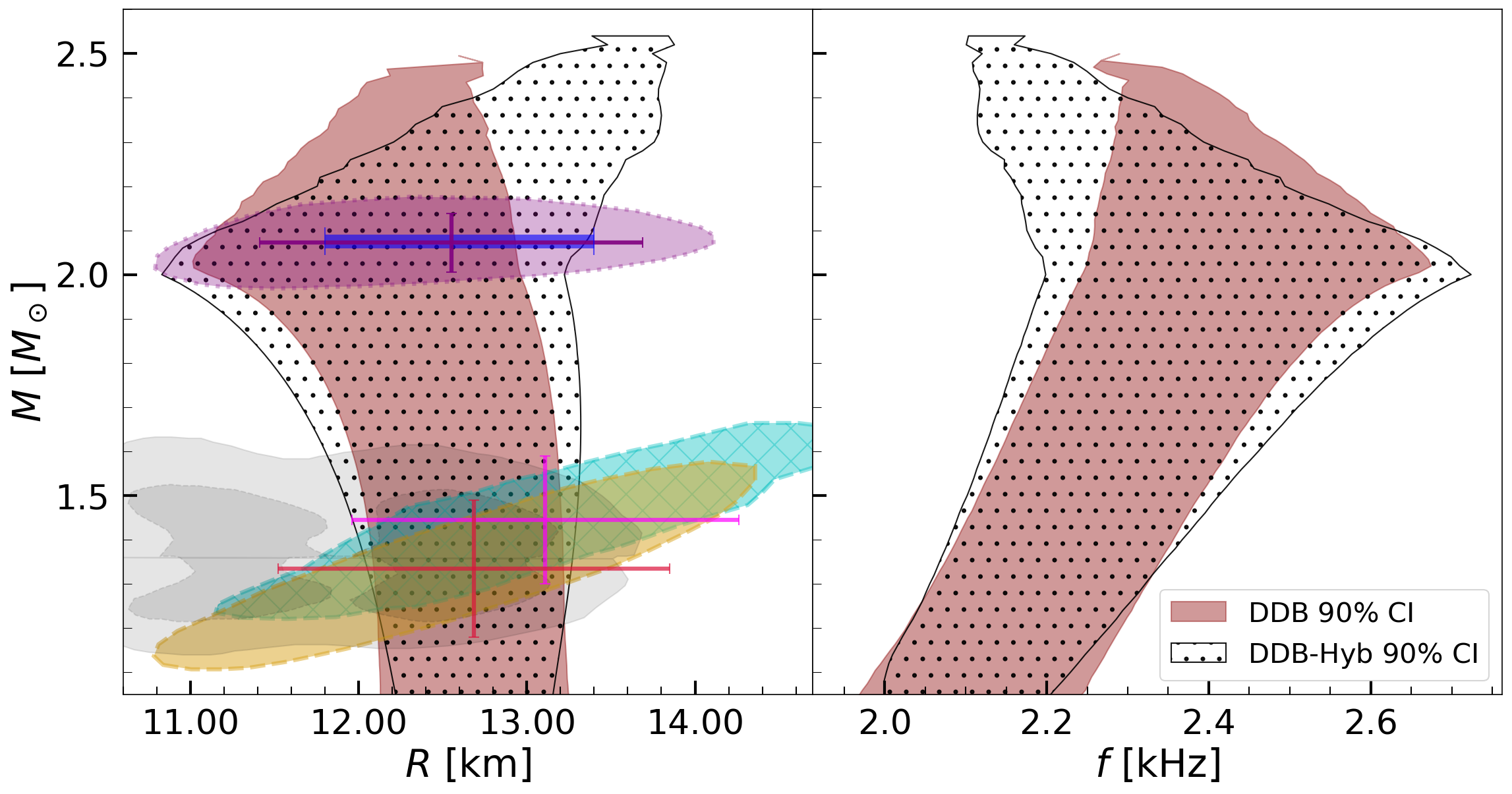}
    \caption{\ac{ns} mass ($M$)-radii ($R$) and $f$ mode frequency-mass ($M$) region obtained from the 90\% CI for the conditional probabilities $P(R|M)$ (left) and $P(f|M)$ (right) for \ac{ddbhyb} (black dotted) and \ac{ddb} (dark red). The blue horizontal bar on the left panel indicates the 90\% CI radius for a 2.08$M_\odot$ star determined in \cite{Miller:2021qha} combining observational data from GW170817 and NICER as well as  nuclear saturation properties. The top and bottom gray regions indicate, respectively, the 90\% (solid) and 50\% (dashed) CI of the LIGO/Virgo analysis for each binary component from the GW170817 event \cite{LIGOScientific:2018hze}. The $1\sigma$ (68\%) credible zone of the 2-D posterior distribution in mass-radii domain from millisecond pulsar PSR J0030+0451 (cyan and yellow) \cite{Riley:2019yda,Miller:2019cac} as well as PSR J0740 + 6620 (violet) \cite{Riley:2021pdl,Miller:2021qha} are shown for the NICER x-rays data. The horizontal (radius) and vertical (mass) error bars reflect the $1\sigma$ credible interval derived for the same NICER data's 1-D marginalized posterior distribution. \label{fig:mrcurve_and_fmode}}
\end{figure}

In Fig.\ref{fig:mrcurve_and_fmode}, we plot the \ac{ns} mass-radii and $f$ mode frequency-mass regions obtained at 90\% CI for the conditional probabilities $P(R|M)$ (left) and $P(f|M)$ (right) from the mass-radius clouds arising from the ensembles of \ac{eos}s of \ac{ddbhyb} (black dotted) and \ac{ddb} (dark red). The blue horizontal bar on the left panel indicates the 90\% CI radius for a 2.08$M_\odot$ star determined in Ref. \cite{Miller:2021qha} combining observational data from GW170817 and NICER as well as  nuclear data. The top and bottom gray regions indicate, the 90\% (solid) and 50\% (dashed) CI of the LIGO/Virgo analysis for each binary component from the GW170817 event \cite{LIGOScientific:2018hze} respectively. The $1\sigma$ (68\%) credible zone of the 2-D posterior distribution in mass-radii domain from millisecond pulsar PSR J0030+0451 (cyan and yellow) \cite{Riley:2019yda,Miller:2019cac} as well as PSR J0740 + 6620 (violet) \cite{Riley:2021pdl,Miller:2021qha} are shown for the NICER x-rays data. The horizontal (radius) and vertical (mass) error bars reflect the $1\sigma$ credible interval derived for the same NICER data's 1-D marginalized posterior distribution. The mass-radius domain for the \ac{ddbhyb} set sweeps a wider  range than the \ac{ddb} set, restricted to nucleonic degrees of freedom. The \ac{ddbhyb} set constrained by \ac{pqcd} at high density leads to larger radii for high mass \ac{ns}. We conclude that the present observational constraints either obtained from GW170817 or NICER cannot rule out the existence of exotic degrees of freedom. In the right panel, we see that the 90\% CI for $P(f|M)$ $f$ mode frequency $f \in \left[1.95,2.7\right]$ kHz for both the \ac{ddb} and \ac{ddbhyb} sets. The range is smaller for low \ac{ns} mass and as the mass increases the 90\% CI for $f$ mode frequency increases. The $f$ mode frequency of a \ac{ns} above 2M$_\odot$ mass is in the range (2.1-2.7) kHz and (2.3-2.65) kHz for the \ac{ddbhyb} and \ac{ddb} sets, respectively. As mentioned in the earlier sections, the solutions for $f$ mode obtained in this work are within the Cowling approximation (neglecting perturbations of the background metric). It was shown that the Cowling approximation can overestimate the quadrupolar $f$ mode frequency of \ac{ns}s by up to 30 to 10 \% for \ac{ns} masses in the range (1.0-2.5) M$_\odot$ compared to the frequency obtained in the linearised \ac{gr} formalism \cite{Benhar:2004xg, Doneva:2013zqa, Pradhan:2022vdf}. The accurate measurement of $f$ modes may further constrain \ac{eos} to a narrower range. Besides, a star of 2$M_{\odot}$ with a low $f$ mode frequency may indicate an existence of non-nucleonic degrees of freedom.
\begin{figure*}[htpb]
    \centering 
    \includegraphics[width=.33\textwidth,angle=0]{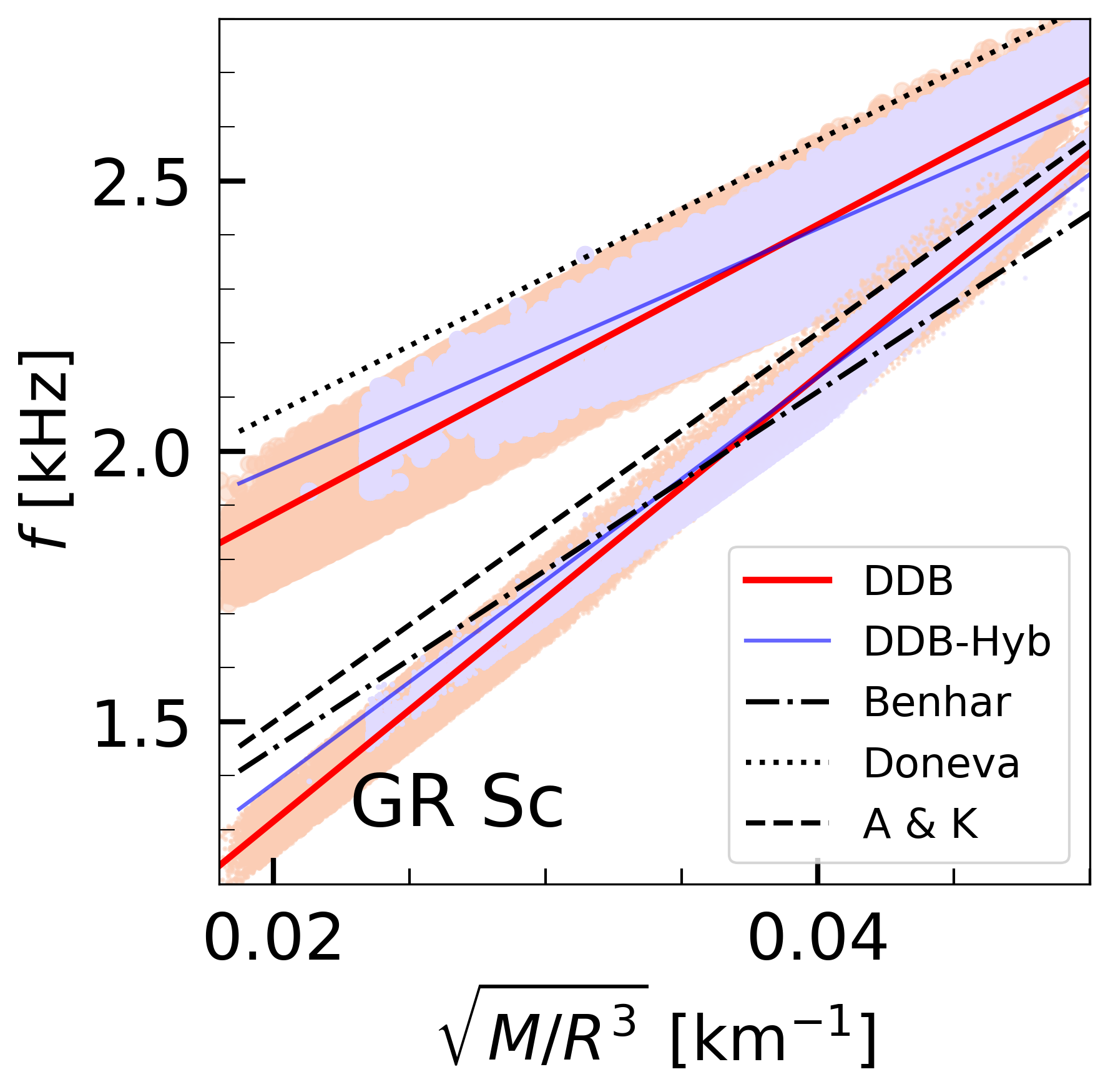} 
    \includegraphics[width=.33\textwidth,angle=0]{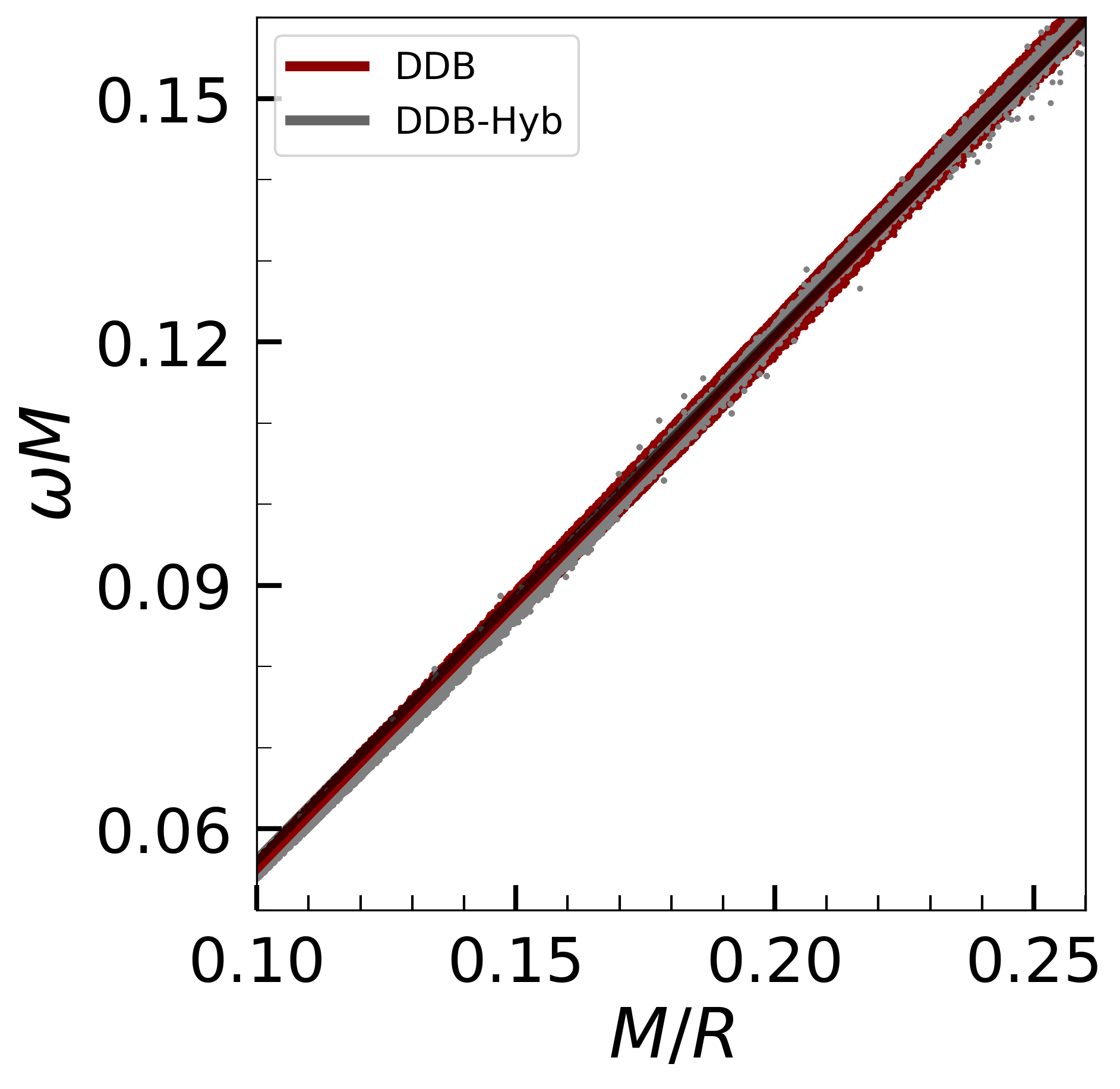}
    \includegraphics[width=.33\textwidth,angle=0]{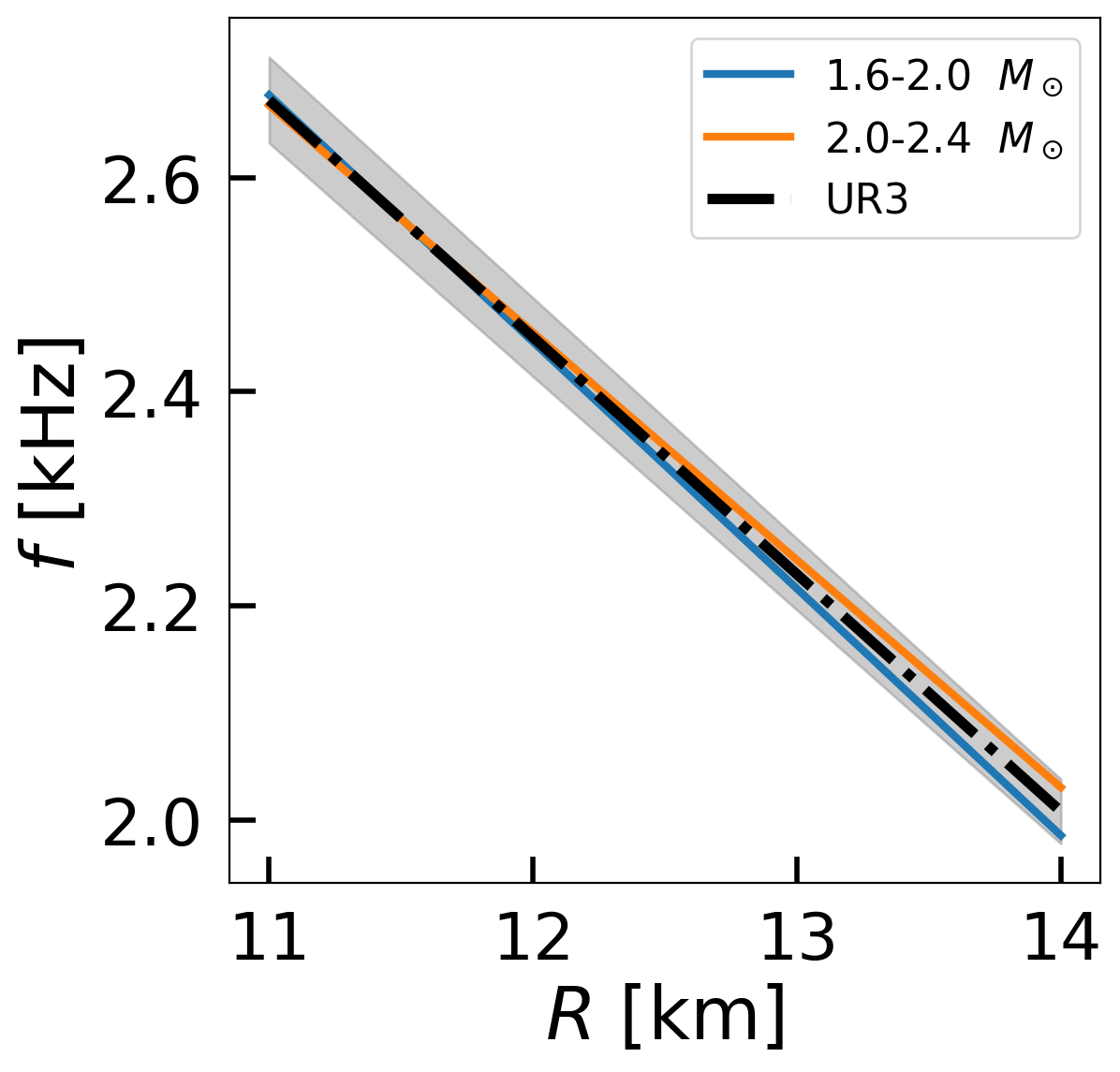}
    \caption{\ac{ur}s obtained with our sets of \ac{eos}s, namely \ac{ddbhyb} and \ac{ddb}. \ac{ur}1 (left): The frequency of the $f$ mode is plotted as a function of the square root of the average density, and corresponding GR scaled data according to \cite{Yoshida:1997bf} (bottom band)  as explained in the text. Also included are results from Doneva {\it et al.} \cite{Doneva:2013zqa} obtained in the Cowling approximation,  Andersson \& Kokkotas \cite{Andersson:1997rn} and Benhar {\it et al.} \cite{Benhar:1998au} calculated with full GR; \ac{ur}2a (center): The universality among $\omega M$ and $M/R$ obtained with both datasets; and \ac{ur}3 (right) the universal linear relations among $f$ mode frequency and radii of \ac{ns} with masses ranging from 1.6 to 2.4 M$_\odot$ in a step of 0.2M$_\odot$. The band corresponds to 90\% CI \label{fig:omem_and_omer}}
\end{figure*}

In Fig. \ref{fig:omem_and_omer}, we have studied two known \ac{ur}s involving the $f$ mode frequency with global properties of \ac{ns}, often studied in literature with a limited set of \ac{eos}s. In particular, we name \ac{ur}1 the UR between the $f$ mode frequency  and the square root of the average star density $\sqrt{M/R^3}$, and \ac{ur}2a the UR involving the $\omega M$ versus the compactness $M/R$, where $\omega=2\pi f$. We have analysed their robustness with our \ac{eos} sets, \ac{ddbhyb} and \ac{ddb}. We have also found a new and direct relation between the $f$ modes frequency, $f$, and radius, $R$, with the help of the existing strong correlation between them. In the left panel of the figure we show \ac{ur}1:
\begin{equation} \label{ur1}
f = a \sqrt{(M/R^3)} + b.
\end{equation}
It has been shown in Refs. \cite{Andersson:1997rn, Kokkotas:1999mn} that the average density can be well parameterized via the $f$ mode frequency. The following values of $a$ and $b$ have been obtained: $a= 22.27\pm0.023~(26.76 \pm 0.01) ~{\rm kHz.km}$, $b= 1.520\pm0.001~ (1.348\pm0.001)$ kHz for \ac{ddbhyb} (\ac{ddb}). The maximum relative percentage error obtained for \ac{ur}1 within 90\% CI is 6.0\%(4.5\%) for \ac{ddbhyb} (\ac{ddb}). We verify that the \ac{ur}1 depends slightly on the  \ac{eos}, reflected in a relative dispersion of $\sim 5\%$ at 90\%CI. Also, the slope of the medians depend on the dataset, with the nucleonic data set DDB presenting a  15\% larger slope, and similar to the one obtained in \cite{Doneva:2013zqa} which was calculated with realistic nucleonic EOS, and is at the upper limit of our 90\% CI. It is important to take note that this particular work has been executed utilizing the Cowling approximations \cite{Sotani:2021nlx, Kumar:2021hzo, McDermott:1983apj} as previously referenced. In the scholarly publication by Yoshida et al. \cite{Yoshida:1997bf}, a comparative analysis was performed  between the outcomes obtained from complete linearized General Relativity (GR) and those acquired through the Cowling approximations. The findings of their study reveal that, for  $l=2$, the $f$ mode is overestimated by 30\% and 15\% when the compactness values of $M/R$ are 0.05 and 0.2, respectively. Using this as a linear relation, we have scaled the solutions obtained in the Cowling approximation, for both \ac{ddb} and \ac{ddbhyb},  see the bottom band in Fig. \ref{fig:omem_and_omer} left panel designated by GR solutions. It is interesting to notice that the scaled frequencies are  compatible with the full GR solutions obtained in the literature. Notice that the dispersion is smaller, but still corresponds to a 5\% relative uncertainty. In Andersson \& Kokkotas (Benhar et al) the authors have obtained the following parameters $a= 35.9 (33.0) {\rm kHz.km}$ and $b=0.78~(0.79)$ kHz \cite{Andersson:1997rn, Kokkotas:1999mn, Benhar:2004xg}, the difference between both works being the \ac{eos} considered in the study. In those studies the linearised \ac{gr} equations were solved, and, as expected, lower frequencies have been determined. In Ref. \cite{Doneva:2013zqa}, the oscillations of non-rotating and fast rotating \ac{ns}s have been explored with a different set of \ac{eos}s based on microscopic theories within the Cowling approximation. The values of the coefficients of the \ac{ur}1 obtained were $a=25.32~{\rm kHz.km}$ and $b= 1.562$ kHz, which are at the 90\% CI upper limit of the relations we have obtained.

In center panel of the Fig. \ref{fig:omem_and_omer} we display \ac{ur}2a: 
\begin{align}
\omega M = a \left(\frac{M}{R}\right) + b \label{uni1}
\end{align} 
obtained for both \ac{ddbhyb} and \ac{ddb} sets, with $a = 0.6474 \pm 4.6\times 10^{-5}$ ($a = 0.6549 \pm 2.6\times 10^{-5}$) and $b = -0.0085 \pm 1.05\times 10^{-5}$ ($b = -0.0103 \pm 6.18\times 10^{-6}$) for \ac{ddbhyb} (\ac{ddb}) set. Both the coefficients are dimensionless. The maximum relative percentage error obtained for \ac{ur}2a within 90\% CI is 3.78\% (2.20\%) for \ac{ddbhyb} (\ac{ddb}) set. The values of the slope and intercept for \ac{ur}2a are also compatible with the ones obtained in Ref. \cite{Pradhan:2020amo} within the Cowling approximation with a few nucleonic and hyperonic \ac{eos}s as $a= 0.65765$ and $b=0.0127866$, respectively. We have also obtained a relation as \ac{ur}2b for $\omega R$ as $\omega R = a \left(\frac{M}{R}\right)^2 + b \left(\frac{M}{R}\right) + c$. The coefficients are found to be $a=-3.0369 \pm 0.0013 (-3.1844 \pm 0.0020)$, $b=1.5829 \pm 0.0005 (1.6288 \pm 0.0008)$ and $c=0.4095 \pm 5\times 10^{-5} (0.4087 \pm 7\times 10^{-5})$ for \ac{ddbhyb} (\ac{ddb}) set, all the coefficients are dimensionless. In this case the maximum relative percentage error is 2.6\%~(1.6\%) in the set \ac{ddbhyb} (\ac{ddb}). Compared with \ac{ur}1, the relative maximum uncertainty is smaller for \ac{ur}2a and \ac{ur}2b for both \ac{ddbhyb} and  \ac{ddb} sets. Using these relations we predict $f$ mode frequencies for the PSR J0740+6620. For this pulsar, the mass and radius are determined as $2.08\pm0.7$ M$_\odot$ and $12.35\pm0.75$ km in \cite{Miller:2021qha} combining observational data from GW170817 and NICER as well as nuclear data. The corresponding mean values of $f$ mode frequency is calculated as 2.35 kHz and 2.36 kHz for \ac{ur}2a and \ac{ur}2b, respectively, with a $\sim 1-4 \%$ intrinsic error in the \ac{ur}s and additional $\sim 10-12 \%$ error due to uncertainty present in mass and radius. A comment regarding the Cowling approximation may be in order. 

We have identified a strong linear correlation between the $f$ mode frequency and \ac{ns} radius $R$ and we are naming it as \ac{ur}3. The values $r\in[0.98,0.99]$ of the Pearson correlation coefficient were obtained between $f$ and $R$ for \ac{ns} with a mass $M\in[1.6,2.4]$ with our two sets of \ac{eos}s. These results can also be traced back from \ac{ur}1 by keeping fixed \ac{ns} mass while noting that the correlation is stronger only for the \ac{ns} of large mass. In the right panel of Fig. \ref{fig:omem_and_omer}, we plot the linear relations between $f$ and $R$. The values of slope $m \in [-0.23, -0.21]$  and intercept are $c \in [5.21,5.0]$ for \ac{ns} mass $M\in[1.6~{\rm to}~2.0,2.0~{\rm to}~2.4] ~M_\odot$. We also plot a marginalized \ac{ur}3 obtained with \ac{ns} masses in the range of 1.6 to 2.4 M$_\odot$ with a slope, ($m=-0.22$) and an intercept, ($c=5.1$). This gives $\approx 1.5\%$ relative residual within 90\% CI. We expect that the correlation is also present if the full \ac{gr} solutions are considered. Taking this correction factor into account, the new relation (\ac{ur}3) will be very useful for the upcoming future detection in order to constrain \ac{ns} radius of massive \ac{ns} precisely. For example, in order to measure a radius of a \ac{ns} with $\sim 0.2$ km uncertainty, the $f$ mode frequency needs to be measured within $\sim 2\%$ uncertainty.

{\it Summary and conclusion - } \label{sec:Summary_and_conclusion}
The \ac{qnm}s are related with the viscous properties of matter. In the future, precise measurements of them can put constraints on the \ac{eos} of dense matter. We have studied the $f$ mode frequency among the \ac{qnm}s, which is in the sensitivity band of the future gravitational waves networks \cite{Pratten:2019sed}. We have calculated the $f$ mode frequency within the Cowling approximation with a nucleonic set of 14,000 \ac{eos}s (\ac{ddb} set), obtained in Ref. \cite{Malik:2022zol} based on the \ac{rmf} theory, constrained by existing observational, theoretical and experimental data through Bayesian analysis. We have also generated an ensemble of \ac{eos}s using \ac{ddb} below twice saturation density ($\rho\leq2\rho_0$) and \ac{pqcd} at high densities ($\rho\geq\rho_0$) as in Ref.\cite{Annala2019}.  Two piecewise polytropes have been used to interpolate the region from $2\rho_0$ to $40\rho_0$. Implementing the constraints of causality and maximum mass $M_{\rm max} \geq 2.0 M_{\odot}$ a set of 55000 \ac{ddbhyb} typed \ac{eos}s has obtained. The mass-radius cloud that we obtain from the ensembles of these \ac{eos}s is consistent with the $\rm{GW}170817$ joint probability distribution as well as the recent NICER observations of mass and radius. We have analyzed the robustness of a few previously known universal relations, UR1 and UR2, and confirmed the robustness of UR2. UR1 shows a dispersion of 5\% relative uncertainty at 90\% CI, and a  15\% smaller slope for the \ac{ddbhyb} compared with the \ac{ddb} set. We also found a novel strong correlation between the $f$ mode frequency, $f$, and the radius, $R$, for a \ac{ns} of mass in the range (1.6-2.4) M$_\odot$. These new direct relations between $f$ and $R$ will allow an accurate determination of the radius of \ac{ns} using future $f$ mode detection.

We show that the quadrupolar $f$ mode frequencies obtained in Cowling approximation of \ac{ns} of masses 2.0M$_\odot$ and above lie in the range (2.1-2.7) kHz and (2.3-2.65) kHz for \ac{ddbhyb} and \ac{ddb} sets, respectively. We use this \ac{ur}s to predict the $f$ mode frequencies of the NICER observations and obtain $\sim$2.35\ (2.0) kHz in Cowling approximation (linearized GR) for the PSR J0740+6620 which interestingly lies within the sensitivity band of the future gravitational wave detector networks \cite{Pratten:2019sed} for the detection of gravitational waves. It was shown that a two solar mass star with a low $f$ mode frequency may indicate the existence of non-nucleonic degrees of freedom. In the future, a detailed investigation of how this frequency is correlated with the individual component of the \ac{eos} or different particle compositions in \ac{ns} core will be carried out.

{\it Acknowledgments-} The authors acknowledge the Laboratory for Advanced Computing at the University of Coimbra for providing {HPC} resources for this research results reported within this paper, URL: \hyperlink{https://www.uc.pt/lca}{https://www.uc.pt/lca}. T.M and C.P would like to thank national funds from FCT (Fundação para a Ciência e a Tecnologia, I.P, Portugal) under Projects No. UID/\-FIS/\-04564/\-2019, No. UIDP/\-04564/\-2020, No. UIDB/\-04564/\-2020, and No. POCI-01-0145-FEDER-029912 with financial support from Science, Technology, and Innovation, in its FEDER component, and by the FCT/MCTES budget through national funds (OE).

\begin{acknowledgments}
\end{acknowledgments}

\bibliographystyle{apsrev4-1}
\bibliography{zzpoly.bib}

\begin{thebibliography}{53}%
\makeatletter
\providecommand \@ifxundefined [1]{%
 \@ifx{#1\undefined}
}%
\providecommand \@ifnum [1]{%
 \ifnum #1\expandafter \@firstoftwo
 \else \expandafter \@secondoftwo
 \fi
}%
\providecommand \@ifx [1]{%
 \ifx #1\expandafter \@firstoftwo
 \else \expandafter \@secondoftwo
 \fi
}%
\providecommand \natexlab [1]{#1}%
\providecommand \enquote  [1]{``#1''}%
\providecommand \bibnamefont  [1]{#1}%
\providecommand \bibfnamefont [1]{#1}%
\providecommand \citenamefont [1]{#1}%
\providecommand \href@noop [0]{\@secondoftwo}%
\providecommand \href [0]{\begingroup \@sanitize@url \@href}%
\providecommand \@href[1]{\@@startlink{#1}\@@href}%
\providecommand \@@href[1]{\endgroup#1\@@endlink}%
\providecommand \@sanitize@url [0]{\catcode `\\12\catcode `\$12\catcode
  `\&12\catcode `\#12\catcode `\^12\catcode `\_12\catcode `\%12\relax}%
\providecommand \@@startlink[1]{}%
\providecommand \@@endlink[0]{}%
\providecommand \url  [0]{\begingroup\@sanitize@url \@url }%
\providecommand \@url [1]{\endgroup\@href {#1}{\urlprefix }}%
\providecommand \urlprefix  [0]{URL }%
\providecommand \Eprint [0]{\href }%
\providecommand \doibase [0]{http://dx.doi.org/}%
\providecommand \selectlanguage [0]{\@gobble}%
\providecommand \bibinfo  [0]{\@secondoftwo}%
\providecommand \bibfield  [0]{\@secondoftwo}%
\providecommand \translation [1]{[#1]}%
\providecommand \BibitemOpen [0]{}%
\providecommand \bibitemStop [0]{}%
\providecommand \bibitemNoStop [0]{.\EOS\space}%
\providecommand \EOS [0]{\spacefactor3000\relax}%
\providecommand \BibitemShut  [1]{\csname bibitem#1\endcsname}%
\let\auto@bib@innerbib\@empty
\bibitem [{\citenamefont {{Glendenning}}(1996)}]{book.Glendenning1996}%
  \BibitemOpen
  \bibfield  {author} {\bibinfo {author} {\bibfnamefont {N.~K.}\ \bibnamefont
  {{Glendenning}}},\ }\href {\doibase 10.1007/978-1-4684-0491-3} {\emph
  {\bibinfo {title} {{Compact Stars}}}}\ (\bibinfo  {publisher} {Springer New
  York, NY},\ \bibinfo {year} {1996})\BibitemShut {NoStop}%
\bibitem [{\citenamefont {{Haensel}}\ \emph {et~al.}(2007)\citenamefont
  {{Haensel}}, \citenamefont {{Potekhin}},\ and\ \citenamefont
  {{Yakovlev}}}]{book.Haensel2007}%
  \BibitemOpen
  \bibfield  {author} {\bibinfo {author} {\bibfnamefont {P.}~\bibnamefont
  {{Haensel}}}, \bibinfo {author} {\bibfnamefont {A.~Y.}\ \bibnamefont
  {{Potekhin}}}, \ and\ \bibinfo {author} {\bibfnamefont {D.~G.}\ \bibnamefont
  {{Yakovlev}}},\ }\href {\doibase 10.1007/978-0-387-47301-7} {\emph {\bibinfo
  {title} {{Neutron Stars 1 : Equation of State and Structure}}}},\ Vol.\
  \bibinfo {volume} {326}\ (\bibinfo  {publisher} {Springer New York, NY},\
  \bibinfo {year} {2007})\BibitemShut {NoStop}%
\bibitem [{\citenamefont {Rezzolla}\ \emph {et~al.}(2018)\citenamefont
  {Rezzolla}, \citenamefont {Pizzochero}, \citenamefont {Jones}, \citenamefont
  {Rea},\ and\ \citenamefont {Vida\~na}}]{Rezzolla:2018jee}%
  \BibitemOpen
  \bibinfo {editor} {\bibfnamefont {L.}~\bibnamefont {Rezzolla}}, \bibinfo
  {editor} {\bibfnamefont {P.}~\bibnamefont {Pizzochero}}, \bibinfo {editor}
  {\bibfnamefont {D.~I.}\ \bibnamefont {Jones}}, \bibinfo {editor}
  {\bibfnamefont {N.}~\bibnamefont {Rea}}, \ and\ \bibinfo {editor}
  {\bibfnamefont {I.}~\bibnamefont {Vida\~na}},\ eds.,\ \href {\doibase
  10.1007/978-3-319-97616-7} {\emph {\bibinfo {title} {{The Physics and
  Astrophysics of Neutron Stars}}}},\ Vol.\ \bibinfo {volume} {457}\ (\bibinfo
  {publisher} {Springer},\ \bibinfo {year} {2018})\BibitemShut {NoStop}%
\bibitem [{\citenamefont
  {Schaffner-Bielich}(2020)}]{Schaffner-Bielich:2020psc}%
  \BibitemOpen
  \bibfield  {author} {\bibinfo {author} {\bibfnamefont {J.}~\bibnamefont
  {Schaffner-Bielich}},\ }\href {\doibase 10.1017/9781316848357} {\emph
  {\bibinfo {title} {{Compact Star Physics}}}}\ (\bibinfo  {publisher}
  {Cambridge University Press},\ \bibinfo {year} {2020})\BibitemShut {NoStop}%
\bibitem [{\citenamefont {Lindblom}\ and\ \citenamefont
  {Indik}(2012)}]{Lindblom2012}%
  \BibitemOpen
  \bibfield  {author} {\bibinfo {author} {\bibfnamefont {L.}~\bibnamefont
  {Lindblom}}\ and\ \bibinfo {author} {\bibfnamefont {N.~M.}\ \bibnamefont
  {Indik}},\ }\href {\doibase 10.1103/PhysRevD.86.084003} {\bibfield  {journal}
  {\bibinfo  {journal} {Phys. Rev. D}\ }\textbf {\bibinfo {volume} {86}},\
  \bibinfo {pages} {084003} (\bibinfo {year} {2012})},\ \Eprint
  {http://arxiv.org/abs/1207.3744} {arXiv:1207.3744 [astro-ph.HE]} \BibitemShut
  {NoStop}%
\bibitem [{\citenamefont {Kurkela}\ \emph {et~al.}(2014)\citenamefont
  {Kurkela}, \citenamefont {Fraga}, \citenamefont {Schaffner-Bielich},\ and\
  \citenamefont {Vuorinen}}]{Kurkela:2014vha}%
  \BibitemOpen
  \bibfield  {author} {\bibinfo {author} {\bibfnamefont {A.}~\bibnamefont
  {Kurkela}}, \bibinfo {author} {\bibfnamefont {E.~S.}\ \bibnamefont {Fraga}},
  \bibinfo {author} {\bibfnamefont {J.}~\bibnamefont {Schaffner-Bielich}}, \
  and\ \bibinfo {author} {\bibfnamefont {A.}~\bibnamefont {Vuorinen}},\ }\href
  {\doibase 10.1088/0004-637X/789/2/127} {\bibfield  {journal} {\bibinfo
  {journal} {Astrophys. J.}\ }\textbf {\bibinfo {volume} {789}},\ \bibinfo
  {pages} {127} (\bibinfo {year} {2014})},\ \Eprint
  {http://arxiv.org/abs/1402.6618} {arXiv:1402.6618 [astro-ph.HE]} \BibitemShut
  {NoStop}%
\bibitem [{\citenamefont {Most}\ \emph {et~al.}(2018)\citenamefont {Most},
  \citenamefont {Weih}, \citenamefont {Rezzolla},\ and\ \citenamefont
  {Schaffner-Bielich}}]{Most:2018hfd}%
  \BibitemOpen
  \bibfield  {author} {\bibinfo {author} {\bibfnamefont {E.~R.}\ \bibnamefont
  {Most}}, \bibinfo {author} {\bibfnamefont {L.~R.}\ \bibnamefont {Weih}},
  \bibinfo {author} {\bibfnamefont {L.}~\bibnamefont {Rezzolla}}, \ and\
  \bibinfo {author} {\bibfnamefont {J.}~\bibnamefont {Schaffner-Bielich}},\
  }\href {\doibase 10.1103/PhysRevLett.120.261103} {\bibfield  {journal}
  {\bibinfo  {journal} {Phys. Rev. Lett.}\ }\textbf {\bibinfo {volume} {120}},\
  \bibinfo {pages} {261103} (\bibinfo {year} {2018})},\ \Eprint
  {http://arxiv.org/abs/1803.00549} {arXiv:1803.00549 [gr-qc]} \BibitemShut
  {NoStop}%
\bibitem [{\citenamefont {Lope~Oter}\ \emph {et~al.}(2019)\citenamefont
  {Lope~Oter}, \citenamefont {Windisch}, \citenamefont {Llanes-Estrada},\ and\
  \citenamefont {Alford}}]{LopeOter:2019pcq}%
  \BibitemOpen
  \bibfield  {author} {\bibinfo {author} {\bibfnamefont {E.}~\bibnamefont
  {Lope~Oter}}, \bibinfo {author} {\bibfnamefont {A.}~\bibnamefont {Windisch}},
  \bibinfo {author} {\bibfnamefont {F.~J.}\ \bibnamefont {Llanes-Estrada}}, \
  and\ \bibinfo {author} {\bibfnamefont {M.}~\bibnamefont {Alford}},\ }\href
  {\doibase 10.1088/1361-6471/ab2567} {\bibfield  {journal} {\bibinfo
  {journal} {J. Phys. G}\ }\textbf {\bibinfo {volume} {46}},\ \bibinfo {pages}
  {084001} (\bibinfo {year} {2019})},\ \Eprint
  {http://arxiv.org/abs/1901.05271} {arXiv:1901.05271 [gr-qc]} \BibitemShut
  {NoStop}%
\bibitem [{\citenamefont {Annala}\ \emph
  {et~al.}(2020{\natexlab{a}})\citenamefont {Annala}, \citenamefont {Gorda},
  \citenamefont {Kurkela}, \citenamefont {N\"attil\"a},\ and\ \citenamefont
  {Vuorinen}}]{Annala2019}%
  \BibitemOpen
  \bibfield  {author} {\bibinfo {author} {\bibfnamefont {E.}~\bibnamefont
  {Annala}}, \bibinfo {author} {\bibfnamefont {T.}~\bibnamefont {Gorda}},
  \bibinfo {author} {\bibfnamefont {A.}~\bibnamefont {Kurkela}}, \bibinfo
  {author} {\bibfnamefont {J.}~\bibnamefont {N\"attil\"a}}, \ and\ \bibinfo
  {author} {\bibfnamefont {A.}~\bibnamefont {Vuorinen}},\ }\href {\doibase
  10.1038/s41567-020-0914-9} {\bibfield  {journal} {\bibinfo  {journal} {Nature
  Phys.}\ }\textbf {\bibinfo {volume} {16}},\ \bibinfo {pages} {907} (\bibinfo
  {year} {2020}{\natexlab{a}})},\ \Eprint {http://arxiv.org/abs/1903.09121}
  {arXiv:1903.09121 [astro-ph.HE]} \BibitemShut {NoStop}%
\bibitem [{\citenamefont {Annala}\ \emph {et~al.}(2022)\citenamefont {Annala},
  \citenamefont {Gorda}, \citenamefont {Katerini}, \citenamefont {Kurkela},
  \citenamefont {N\"attil\"a}, \citenamefont {Paschalidis},\ and\ \citenamefont
  {Vuorinen}}]{Annala:2021gom}%
  \BibitemOpen
  \bibfield  {author} {\bibinfo {author} {\bibfnamefont {E.}~\bibnamefont
  {Annala}}, \bibinfo {author} {\bibfnamefont {T.}~\bibnamefont {Gorda}},
  \bibinfo {author} {\bibfnamefont {E.}~\bibnamefont {Katerini}}, \bibinfo
  {author} {\bibfnamefont {A.}~\bibnamefont {Kurkela}}, \bibinfo {author}
  {\bibfnamefont {J.}~\bibnamefont {N\"attil\"a}}, \bibinfo {author}
  {\bibfnamefont {V.}~\bibnamefont {Paschalidis}}, \ and\ \bibinfo {author}
  {\bibfnamefont {A.}~\bibnamefont {Vuorinen}},\ }\href {\doibase
  10.1103/PhysRevX.12.011058} {\bibfield  {journal} {\bibinfo  {journal} {Phys.
  Rev. X}\ }\textbf {\bibinfo {volume} {12}},\ \bibinfo {pages} {011058}
  (\bibinfo {year} {2022})},\ \Eprint {http://arxiv.org/abs/2105.05132}
  {arXiv:2105.05132 [astro-ph.HE]} \BibitemShut {NoStop}%
\bibitem [{\citenamefont {Altiparmak}\ \emph {et~al.}(2022)\citenamefont
  {Altiparmak}, \citenamefont {Ecker},\ and\ \citenamefont
  {Rezzolla}}]{Altiparmak:2022bke}%
  \BibitemOpen
  \bibfield  {author} {\bibinfo {author} {\bibfnamefont {S.}~\bibnamefont
  {Altiparmak}}, \bibinfo {author} {\bibfnamefont {C.}~\bibnamefont {Ecker}}, \
  and\ \bibinfo {author} {\bibfnamefont {L.}~\bibnamefont {Rezzolla}},\ }\href
  {\doibase 10.3847/2041-8213/ac9b2a} {\bibfield  {journal} {\bibinfo
  {journal} {Astrophys. J. Lett.}\ }\textbf {\bibinfo {volume} {939}},\
  \bibinfo {pages} {L34} (\bibinfo {year} {2022})},\ \Eprint
  {http://arxiv.org/abs/2203.14974} {arXiv:2203.14974 [astro-ph.HE]}
  \BibitemShut {NoStop}%
\bibitem [{\citenamefont {de~Tovar}\ \emph {et~al.}(2021)\citenamefont
  {de~Tovar}, \citenamefont {Ferreira},\ and\ \citenamefont
  {Provid\^encia}}]{Tovar2021}%
  \BibitemOpen
  \bibfield  {author} {\bibinfo {author} {\bibfnamefont {P.~B.}\ \bibnamefont
  {de~Tovar}}, \bibinfo {author} {\bibfnamefont {M.}~\bibnamefont {Ferreira}},
  \ and\ \bibinfo {author} {\bibfnamefont {C.}~\bibnamefont {Provid\^encia}},\
  }\href {\doibase 10.1103/PhysRevD.104.123036} {\bibfield  {journal} {\bibinfo
   {journal} {Phys. Rev. D}\ }\textbf {\bibinfo {volume} {104}},\ \bibinfo
  {pages} {123036} (\bibinfo {year} {2021})},\ \Eprint
  {http://arxiv.org/abs/2112.05551} {arXiv:2112.05551 [nucl-th]} \BibitemShut
  {NoStop}%
\bibitem [{\citenamefont {Imam}\ \emph {et~al.}(2022)\citenamefont {Imam},
  \citenamefont {Patra}, \citenamefont {Mondal}, \citenamefont {Malik},\ and\
  \citenamefont {Agrawal}}]{Imam2021}%
  \BibitemOpen
  \bibfield  {author} {\bibinfo {author} {\bibfnamefont {S.~M.~A.}\
  \bibnamefont {Imam}}, \bibinfo {author} {\bibfnamefont {N.~K.}\ \bibnamefont
  {Patra}}, \bibinfo {author} {\bibfnamefont {C.}~\bibnamefont {Mondal}},
  \bibinfo {author} {\bibfnamefont {T.}~\bibnamefont {Malik}}, \ and\ \bibinfo
  {author} {\bibfnamefont {B.~K.}\ \bibnamefont {Agrawal}},\ }\href {\doibase
  10.1103/PhysRevC.105.015806} {\bibfield  {journal} {\bibinfo  {journal}
  {Phys. Rev. C}\ }\textbf {\bibinfo {volume} {105}},\ \bibinfo {pages}
  {015806} (\bibinfo {year} {2022})},\ \Eprint
  {http://arxiv.org/abs/2110.15776} {arXiv:2110.15776 [nucl-th]} \BibitemShut
  {NoStop}%
\bibitem [{\citenamefont {Mondal}\ and\ \citenamefont
  {Gulminelli}(2022)}]{Mondal2021}%
  \BibitemOpen
  \bibfield  {author} {\bibinfo {author} {\bibfnamefont {C.}~\bibnamefont
  {Mondal}}\ and\ \bibinfo {author} {\bibfnamefont {F.}~\bibnamefont
  {Gulminelli}},\ }\href {\doibase 10.1103/PhysRevD.105.083016} {\bibfield
  {journal} {\bibinfo  {journal} {Phys. Rev. D}\ }\textbf {\bibinfo {volume}
  {105}},\ \bibinfo {pages} {083016} (\bibinfo {year} {2022})},\ \Eprint
  {http://arxiv.org/abs/2111.04520} {arXiv:2111.04520 [nucl-th]} \BibitemShut
  {NoStop}%
\bibitem [{\citenamefont {Essick}\ \emph {et~al.}(2021)\citenamefont {Essick},
  \citenamefont {Tews}, \citenamefont {Landry},\ and\ \citenamefont
  {Schwenk}}]{Essick:2021vlx}%
  \BibitemOpen
  \bibfield  {author} {\bibinfo {author} {\bibfnamefont {R.}~\bibnamefont
  {Essick}}, \bibinfo {author} {\bibfnamefont {I.}~\bibnamefont {Tews}},
  \bibinfo {author} {\bibfnamefont {P.}~\bibnamefont {Landry}}, \ and\ \bibinfo
  {author} {\bibfnamefont {A.}~\bibnamefont {Schwenk}},\ }\href {\doibase
  10.1103/PhysRevLett.127.192701} {\bibfield  {journal} {\bibinfo  {journal}
  {Phys. Rev. Lett.}\ }\textbf {\bibinfo {volume} {127}},\ \bibinfo {pages}
  {192701} (\bibinfo {year} {2021})}\BibitemShut {NoStop}%
\bibitem [{\citenamefont {Chirenti}\ \emph {et~al.}(2012)\citenamefont
  {Chirenti}, \citenamefont {Silveira},\ and\ \citenamefont
  {Aguiar}}]{Chirenti:2012wn}%
  \BibitemOpen
  \bibfield  {author} {\bibinfo {author} {\bibfnamefont {C.}~\bibnamefont
  {Chirenti}}, \bibinfo {author} {\bibfnamefont {P.~R.}\ \bibnamefont
  {Silveira}}, \ and\ \bibinfo {author} {\bibfnamefont {O.~D.}\ \bibnamefont
  {Aguiar}},\ }\href {\doibase 10.1142/S2010194512008185} {\bibfield  {journal}
  {\bibinfo  {journal} {Int. J. Mod. Phys. Conf. Ser.}\ }\textbf {\bibinfo
  {volume} {18}},\ \bibinfo {pages} {48} (\bibinfo {year} {2012})},\ \Eprint
  {http://arxiv.org/abs/1205.2001} {arXiv:1205.2001 [gr-qc]} \BibitemShut
  {NoStop}%
\bibitem [{\citenamefont {Pratten}\ \emph {et~al.}(2020)\citenamefont
  {Pratten}, \citenamefont {Schmidt},\ and\ \citenamefont
  {Hinderer}}]{Pratten:2019sed}%
  \BibitemOpen
  \bibfield  {author} {\bibinfo {author} {\bibfnamefont {G.}~\bibnamefont
  {Pratten}}, \bibinfo {author} {\bibfnamefont {P.}~\bibnamefont {Schmidt}}, \
  and\ \bibinfo {author} {\bibfnamefont {T.}~\bibnamefont {Hinderer}},\ }\href
  {\doibase 10.1038/s41467-020-15984-5} {\bibfield  {journal} {\bibinfo
  {journal} {Nature Commun.}\ }\textbf {\bibinfo {volume} {11}},\ \bibinfo
  {pages} {2553} (\bibinfo {year} {2020})},\ \Eprint
  {http://arxiv.org/abs/1905.00817} {arXiv:1905.00817 [gr-qc]} \BibitemShut
  {NoStop}%
\bibitem [{\citenamefont {Andersson}\ and\ \citenamefont
  {Kokkotas}(1996)}]{Andersson:1996pn}%
  \BibitemOpen
  \bibfield  {author} {\bibinfo {author} {\bibfnamefont {N.}~\bibnamefont
  {Andersson}}\ and\ \bibinfo {author} {\bibfnamefont {K.~D.}\ \bibnamefont
  {Kokkotas}},\ }\href {\doibase 10.1103/PhysRevLett.77.4134} {\bibfield
  {journal} {\bibinfo  {journal} {Phys. Rev. Lett.}\ }\textbf {\bibinfo
  {volume} {77}},\ \bibinfo {pages} {4134} (\bibinfo {year} {1996})},\ \Eprint
  {http://arxiv.org/abs/gr-qc/9610035} {arXiv:gr-qc/9610035} \BibitemShut
  {NoStop}%
\bibitem [{\citenamefont {Andersson}\ and\ \citenamefont
  {Kokkotas}(1998)}]{Andersson:1997rn}%
  \BibitemOpen
  \bibfield  {author} {\bibinfo {author} {\bibfnamefont {N.}~\bibnamefont
  {Andersson}}\ and\ \bibinfo {author} {\bibfnamefont {K.~D.}\ \bibnamefont
  {Kokkotas}},\ }\href {\doibase 10.1046/j.1365-8711.1998.01840.x} {\bibfield
  {journal} {\bibinfo  {journal} {Mon. Not. Roy. Astron. Soc.}\ }\textbf
  {\bibinfo {volume} {299}},\ \bibinfo {pages} {1059} (\bibinfo {year}
  {1998})},\ \Eprint {http://arxiv.org/abs/gr-qc/9711088} {arXiv:gr-qc/9711088}
  \BibitemShut {NoStop}%
\bibitem [{\citenamefont {Benhar}\ \emph {et~al.}(1999)\citenamefont {Benhar},
  \citenamefont {Berti},\ and\ \citenamefont {Ferrari}}]{Benhar:1998au}%
  \BibitemOpen
  \bibfield  {author} {\bibinfo {author} {\bibfnamefont {O.}~\bibnamefont
  {Benhar}}, \bibinfo {author} {\bibfnamefont {E.}~\bibnamefont {Berti}}, \
  and\ \bibinfo {author} {\bibfnamefont {V.}~\bibnamefont {Ferrari}},\ }\href
  {\doibase 10.1046/j.1365-8711.1999.02983.x} {\bibfield  {journal} {\bibinfo
  {journal} {Mon. Not. Roy. Astron. Soc.}\ }\textbf {\bibinfo {volume} {310}},\
  \bibinfo {pages} {797} (\bibinfo {year} {1999})},\ \Eprint
  {http://arxiv.org/abs/gr-qc/9901037} {arXiv:gr-qc/9901037} \BibitemShut
  {NoStop}%
\bibitem [{\citenamefont {Benhar}\ \emph {et~al.}(2004)\citenamefont {Benhar},
  \citenamefont {Ferrari},\ and\ \citenamefont {Gualtieri}}]{Benhar:2004xg}%
  \BibitemOpen
  \bibfield  {author} {\bibinfo {author} {\bibfnamefont {O.}~\bibnamefont
  {Benhar}}, \bibinfo {author} {\bibfnamefont {V.}~\bibnamefont {Ferrari}}, \
  and\ \bibinfo {author} {\bibfnamefont {L.}~\bibnamefont {Gualtieri}},\ }\href
  {\doibase 10.1103/PhysRevD.70.124015} {\bibfield  {journal} {\bibinfo
  {journal} {Phys. Rev. D}\ }\textbf {\bibinfo {volume} {70}},\ \bibinfo
  {pages} {124015} (\bibinfo {year} {2004})},\ \Eprint
  {http://arxiv.org/abs/astro-ph/0407529} {arXiv:astro-ph/0407529} \BibitemShut
  {NoStop}%
\bibitem [{\citenamefont {Tsui}\ and\ \citenamefont
  {Leung}(2005)}]{Tsui:2004qd}%
  \BibitemOpen
  \bibfield  {author} {\bibinfo {author} {\bibfnamefont {L.~K.}\ \bibnamefont
  {Tsui}}\ and\ \bibinfo {author} {\bibfnamefont {P.~T.}\ \bibnamefont
  {Leung}},\ }\href {\doibase 10.1111/j.1365-2966.2005.08710.x} {\bibfield
  {journal} {\bibinfo  {journal} {Mon. Not. Roy. Astron. Soc.}\ }\textbf
  {\bibinfo {volume} {357}},\ \bibinfo {pages} {1029} (\bibinfo {year}
  {2005})},\ \Eprint {http://arxiv.org/abs/gr-qc/0412024} {arXiv:gr-qc/0412024}
  \BibitemShut {NoStop}%
\bibitem [{\citenamefont {Chan}\ \emph {et~al.}(2014)\citenamefont {Chan},
  \citenamefont {Sham}, \citenamefont {Leung},\ and\ \citenamefont
  {Lin}}]{Chan:2014kua}%
  \BibitemOpen
  \bibfield  {author} {\bibinfo {author} {\bibfnamefont {T.~K.}\ \bibnamefont
  {Chan}}, \bibinfo {author} {\bibfnamefont {Y.~H.}\ \bibnamefont {Sham}},
  \bibinfo {author} {\bibfnamefont {P.~T.}\ \bibnamefont {Leung}}, \ and\
  \bibinfo {author} {\bibfnamefont {L.~M.}\ \bibnamefont {Lin}},\ }\href
  {\doibase 10.1103/PhysRevD.90.124023} {\bibfield  {journal} {\bibinfo
  {journal} {Phys. Rev. D}\ }\textbf {\bibinfo {volume} {90}},\ \bibinfo
  {pages} {124023} (\bibinfo {year} {2014})},\ \Eprint
  {http://arxiv.org/abs/1408.3789} {arXiv:1408.3789 [gr-qc]} \BibitemShut
  {NoStop}%
\bibitem [{\citenamefont {Sotani}(2021)}]{Sotani:2021nlx}%
  \BibitemOpen
  \bibfield  {author} {\bibinfo {author} {\bibfnamefont {H.}~\bibnamefont
  {Sotani}},\ }\href {\doibase 10.1103/PhysRevD.103.123015} {\bibfield
  {journal} {\bibinfo  {journal} {Phys. Rev. D}\ }\textbf {\bibinfo {volume}
  {103}},\ \bibinfo {pages} {123015} (\bibinfo {year} {2021})},\ \Eprint
  {http://arxiv.org/abs/2105.13212} {arXiv:2105.13212 [astro-ph.HE]}
  \BibitemShut {NoStop}%
\bibitem [{\citenamefont {Sotani}\ and\ \citenamefont
  {Kumar}(2021)}]{Sotani:2021kiw}%
  \BibitemOpen
  \bibfield  {author} {\bibinfo {author} {\bibfnamefont {H.}~\bibnamefont
  {Sotani}}\ and\ \bibinfo {author} {\bibfnamefont {B.}~\bibnamefont {Kumar}},\
  }\href {\doibase 10.1103/PhysRevD.104.123002} {\bibfield  {journal} {\bibinfo
   {journal} {Phys. Rev. D}\ }\textbf {\bibinfo {volume} {104}},\ \bibinfo
  {pages} {123002} (\bibinfo {year} {2021})},\ \Eprint
  {http://arxiv.org/abs/2109.08145} {arXiv:2109.08145 [gr-qc]} \BibitemShut
  {NoStop}%
\bibitem [{\citenamefont {{McDermott}}\ \emph {et~al.}(1983)\citenamefont
  {{McDermott}}, \citenamefont {{van Horn}},\ and\ \citenamefont
  {{Scholl}}}]{McDermott:1983apj}%
  \BibitemOpen
  \bibfield  {author} {\bibinfo {author} {\bibfnamefont {P.~N.}\ \bibnamefont
  {{McDermott}}}, \bibinfo {author} {\bibfnamefont {H.~M.}\ \bibnamefont {{van
  Horn}}}, \ and\ \bibinfo {author} {\bibfnamefont {J.~F.}\ \bibnamefont
  {{Scholl}}},\ }\href {\doibase 10.1086/161006} {\bibfield  {journal}
  {\bibinfo  {journal} {\apj}\ }\textbf {\bibinfo {volume} {268}},\ \bibinfo
  {pages} {837} (\bibinfo {year} {1983})}\BibitemShut {NoStop}%
\bibitem [{\citenamefont {Yoshida}\ and\ \citenamefont
  {Lee}(2002)}]{Yoshida:2002vd}%
  \BibitemOpen
  \bibfield  {author} {\bibinfo {author} {\bibfnamefont {S.}~\bibnamefont
  {Yoshida}}\ and\ \bibinfo {author} {\bibfnamefont {U.}~\bibnamefont {Lee}},\
  }\href {\doibase 10.1051/0004-6361:20021270} {\bibfield  {journal} {\bibinfo
  {journal} {Astron. Astrophys.}\ }\textbf {\bibinfo {volume} {395}},\ \bibinfo
  {pages} {201} (\bibinfo {year} {2002})},\ \Eprint
  {http://arxiv.org/abs/astro-ph/0210591} {arXiv:astro-ph/0210591} \BibitemShut
  {NoStop}%
\bibitem [{\citenamefont {Lau}\ \emph {et~al.}(2019)\citenamefont {Lau},
  \citenamefont {Leung},\ and\ \citenamefont {Lin}}]{Lau:2018mae}%
  \BibitemOpen
  \bibfield  {author} {\bibinfo {author} {\bibfnamefont {S.~Y.}\ \bibnamefont
  {Lau}}, \bibinfo {author} {\bibfnamefont {P.~T.}\ \bibnamefont {Leung}}, \
  and\ \bibinfo {author} {\bibfnamefont {L.~M.}\ \bibnamefont {Lin}},\ }\href
  {\doibase 10.1103/PhysRevD.99.023018} {\bibfield  {journal} {\bibinfo
  {journal} {Phys. Rev. D}\ }\textbf {\bibinfo {volume} {99}},\ \bibinfo
  {pages} {023018} (\bibinfo {year} {2019})},\ \Eprint
  {http://arxiv.org/abs/1808.08107} {arXiv:1808.08107 [astro-ph.HE]}
  \BibitemShut {NoStop}%
\bibitem [{\citenamefont {Bandyopadhyay}\ \emph {et~al.}(2018)\citenamefont
  {Bandyopadhyay}, \citenamefont {Bhat}, \citenamefont {Char},\ and\
  \citenamefont {Chatterjee}}]{Bandyopadhyay:2017dvi}%
  \BibitemOpen
  \bibfield  {author} {\bibinfo {author} {\bibfnamefont {D.}~\bibnamefont
  {Bandyopadhyay}}, \bibinfo {author} {\bibfnamefont {S.~A.}\ \bibnamefont
  {Bhat}}, \bibinfo {author} {\bibfnamefont {P.}~\bibnamefont {Char}}, \ and\
  \bibinfo {author} {\bibfnamefont {D.}~\bibnamefont {Chatterjee}},\ }\href
  {\doibase 10.1140/epja/i2018-12456-y} {\bibfield  {journal} {\bibinfo
  {journal} {Eur. Phys. J. A}\ }\textbf {\bibinfo {volume} {54}},\ \bibinfo
  {pages} {26} (\bibinfo {year} {2018})},\ \Eprint
  {http://arxiv.org/abs/1712.01715} {arXiv:1712.01715 [astro-ph.HE]}
  \BibitemShut {NoStop}%
\bibitem [{\citenamefont {Han}\ and\ \citenamefont
  {Steiner}(2019)}]{Han:2018mtj}%
  \BibitemOpen
  \bibfield  {author} {\bibinfo {author} {\bibfnamefont {S.}~\bibnamefont
  {Han}}\ and\ \bibinfo {author} {\bibfnamefont {A.~W.}\ \bibnamefont
  {Steiner}},\ }\href {\doibase 10.1103/PhysRevD.99.083014} {\bibfield
  {journal} {\bibinfo  {journal} {Phys. Rev. D}\ }\textbf {\bibinfo {volume}
  {99}},\ \bibinfo {pages} {083014} (\bibinfo {year} {2019})},\ \Eprint
  {http://arxiv.org/abs/1810.10967} {arXiv:1810.10967 [nucl-th]} \BibitemShut
  {NoStop}%
\bibitem [{\citenamefont {von Doetinchem}\ \emph {et~al.}(2020)\citenamefont
  {von Doetinchem} \emph {et~al.}}]{vonDoetinchem:2020vbj}%
  \BibitemOpen
  \bibfield  {author} {\bibinfo {author} {\bibfnamefont {P.}~\bibnamefont {von
  Doetinchem}} \emph {et~al.},\ }\href {\doibase 10.1088/1475-7516/2020/08/035}
  {\bibfield  {journal} {\bibinfo  {journal} {JCAP}\ }\textbf {\bibinfo
  {volume} {08}},\ \bibinfo {pages} {035} (\bibinfo {year} {2020})},\ \Eprint
  {http://arxiv.org/abs/2002.04163} {arXiv:2002.04163 [astro-ph.HE]}
  \BibitemShut {NoStop}%
\bibitem [{\citenamefont {Baym}\ \emph {et~al.}(1971)\citenamefont {Baym},
  \citenamefont {Pethick},\ and\ \citenamefont {Sutherland}}]{Baym:1971pw}%
  \BibitemOpen
  \bibfield  {author} {\bibinfo {author} {\bibfnamefont {G.}~\bibnamefont
  {Baym}}, \bibinfo {author} {\bibfnamefont {C.}~\bibnamefont {Pethick}}, \
  and\ \bibinfo {author} {\bibfnamefont {P.}~\bibnamefont {Sutherland}},\
  }\href {\doibase 10.1086/151216} {\bibfield  {journal} {\bibinfo  {journal}
  {Astrophys. J.}\ }\textbf {\bibinfo {volume} {170}},\ \bibinfo {pages} {299}
  (\bibinfo {year} {1971})}\BibitemShut {NoStop}%
\bibitem [{\citenamefont {Carriere}\ \emph {et~al.}(2003)\citenamefont
  {Carriere}, \citenamefont {Horowitz},\ and\ \citenamefont
  {Piekarewicz}}]{Carriere:2002bx}%
  \BibitemOpen
  \bibfield  {author} {\bibinfo {author} {\bibfnamefont {J.}~\bibnamefont
  {Carriere}}, \bibinfo {author} {\bibfnamefont {C.~J.}\ \bibnamefont
  {Horowitz}}, \ and\ \bibinfo {author} {\bibfnamefont {J.}~\bibnamefont
  {Piekarewicz}},\ }\href {\doibase 10.1086/376515} {\bibfield  {journal}
  {\bibinfo  {journal} {Astrophys. J.}\ }\textbf {\bibinfo {volume} {593}},\
  \bibinfo {pages} {463} (\bibinfo {year} {2003})},\ \Eprint
  {http://arxiv.org/abs/nucl-th/0211015} {arXiv:nucl-th/0211015} \BibitemShut
  {NoStop}%
\bibitem [{\citenamefont {Malik}\ \emph {et~al.}(2022)\citenamefont {Malik},
  \citenamefont {Ferreira}, \citenamefont {Agrawal},\ and\ \citenamefont
  {Provid\^encia}}]{Malik:2022zol}%
  \BibitemOpen
  \bibfield  {author} {\bibinfo {author} {\bibfnamefont {T.}~\bibnamefont
  {Malik}}, \bibinfo {author} {\bibfnamefont {M.}~\bibnamefont {Ferreira}},
  \bibinfo {author} {\bibfnamefont {B.~K.}\ \bibnamefont {Agrawal}}, \ and\
  \bibinfo {author} {\bibfnamefont {C.}~\bibnamefont {Provid\^encia}},\ }\href
  {\doibase 10.3847/1538-4357/ac5d3c} {\bibfield  {journal} {\bibinfo
  {journal} {Astrophys. J.}\ }\textbf {\bibinfo {volume} {930}},\ \bibinfo
  {pages} {17} (\bibinfo {year} {2022})},\ \Eprint
  {http://arxiv.org/abs/2201.12552} {arXiv:2201.12552 [nucl-th]} \BibitemShut
  {NoStop}%
\bibitem [{\citenamefont {Tews}\ \emph {et~al.}(2013)\citenamefont {Tews},
  \citenamefont {Kr\"uger}, \citenamefont {Hebeler},\ and\ \citenamefont
  {Schwenk}}]{Tews:2012fj}%
  \BibitemOpen
  \bibfield  {author} {\bibinfo {author} {\bibfnamefont {I.}~\bibnamefont
  {Tews}}, \bibinfo {author} {\bibfnamefont {T.}~\bibnamefont {Kr\"uger}},
  \bibinfo {author} {\bibfnamefont {K.}~\bibnamefont {Hebeler}}, \ and\
  \bibinfo {author} {\bibfnamefont {A.}~\bibnamefont {Schwenk}},\ }\href
  {\doibase 10.1103/PhysRevLett.110.032504} {\bibfield  {journal} {\bibinfo
  {journal} {Phys. Rev. Lett.}\ }\textbf {\bibinfo {volume} {110}},\ \bibinfo
  {pages} {032504} (\bibinfo {year} {2013})},\ \Eprint
  {http://arxiv.org/abs/1206.0025} {arXiv:1206.0025 [nucl-th]} \BibitemShut
  {NoStop}%
\bibitem [{\citenamefont {Hebeler}\ \emph {et~al.}(2013)\citenamefont
  {Hebeler}, \citenamefont {Lattimer}, \citenamefont {Pethick},\ and\
  \citenamefont {Schwenk}}]{Hebeler:2013nza}%
  \BibitemOpen
  \bibfield  {author} {\bibinfo {author} {\bibfnamefont {K.}~\bibnamefont
  {Hebeler}}, \bibinfo {author} {\bibfnamefont {J.~M.}\ \bibnamefont
  {Lattimer}}, \bibinfo {author} {\bibfnamefont {C.~J.}\ \bibnamefont
  {Pethick}}, \ and\ \bibinfo {author} {\bibfnamefont {A.}~\bibnamefont
  {Schwenk}},\ }\href {\doibase 10.1088/0004-637X/773/1/11} {\bibfield
  {journal} {\bibinfo  {journal} {Astrophys. J.}\ }\textbf {\bibinfo {volume}
  {773}},\ \bibinfo {pages} {11} (\bibinfo {year} {2013})},\ \Eprint
  {http://arxiv.org/abs/1303.4662} {arXiv:1303.4662 [astro-ph.SR]} \BibitemShut
  {NoStop}%
\bibitem [{\citenamefont {Kurkela}\ \emph {et~al.}(2010)\citenamefont
  {Kurkela}, \citenamefont {Romatschke},\ and\ \citenamefont
  {Vuorinen}}]{Kurkela:2009gj}%
  \BibitemOpen
  \bibfield  {author} {\bibinfo {author} {\bibfnamefont {A.}~\bibnamefont
  {Kurkela}}, \bibinfo {author} {\bibfnamefont {P.}~\bibnamefont {Romatschke}},
  \ and\ \bibinfo {author} {\bibfnamefont {A.}~\bibnamefont {Vuorinen}},\
  }\href {\doibase 10.1103/PhysRevD.81.105021} {\bibfield  {journal} {\bibinfo
  {journal} {Phys. Rev. D}\ }\textbf {\bibinfo {volume} {81}},\ \bibinfo
  {pages} {105021} (\bibinfo {year} {2010})},\ \Eprint
  {http://arxiv.org/abs/0912.1856} {arXiv:0912.1856 [hep-ph]} \BibitemShut
  {NoStop}%
\bibitem [{\citenamefont {Fraga}\ \emph {et~al.}(2014)\citenamefont {Fraga},
  \citenamefont {Kurkela},\ and\ \citenamefont {Vuorinen}}]{Fraga:2013qra}%
  \BibitemOpen
  \bibfield  {author} {\bibinfo {author} {\bibfnamefont {E.~S.}\ \bibnamefont
  {Fraga}}, \bibinfo {author} {\bibfnamefont {A.}~\bibnamefont {Kurkela}}, \
  and\ \bibinfo {author} {\bibfnamefont {A.}~\bibnamefont {Vuorinen}},\ }\href
  {\doibase 10.1088/2041-8205/781/2/L25} {\bibfield  {journal} {\bibinfo
  {journal} {Astrophys. J. Lett.}\ }\textbf {\bibinfo {volume} {781}},\
  \bibinfo {pages} {L25} (\bibinfo {year} {2014})},\ \Eprint
  {http://arxiv.org/abs/1311.5154} {arXiv:1311.5154 [nucl-th]} \BibitemShut
  {NoStop}%
\bibitem [{\citenamefont {Florian}(1992)}]{FLORIAN1992123}%
  \BibitemOpen
  \bibfield  {author} {\bibinfo {author} {\bibfnamefont {A.}~\bibnamefont
  {Florian}},\ }\href {\doibase https://doi.org/10.1016/0266-8920(92)90015-A}
  {\bibfield  {journal} {\bibinfo  {journal} {Probabilistic Engineering
  Mechanics}\ }\textbf {\bibinfo {volume} {7}},\ \bibinfo {pages} {123}
  (\bibinfo {year} {1992})}\BibitemShut {NoStop}%
\bibitem [{\citenamefont {Lindblom}(2010)}]{Lindblom:2010bb}%
  \BibitemOpen
  \bibfield  {author} {\bibinfo {author} {\bibfnamefont {L.}~\bibnamefont
  {Lindblom}},\ }\href {\doibase 10.1103/PhysRevD.82.103011} {\bibfield
  {journal} {\bibinfo  {journal} {Phys. Rev. D}\ }\textbf {\bibinfo {volume}
  {82}},\ \bibinfo {pages} {103011} (\bibinfo {year} {2010})},\ \Eprint
  {http://arxiv.org/abs/1009.0738} {arXiv:1009.0738 [astro-ph.HE]} \BibitemShut
  {NoStop}%
\bibitem [{\citenamefont {Lindblom}(2018)}]{Lindblom:2018rfr}%
  \BibitemOpen
  \bibfield  {author} {\bibinfo {author} {\bibfnamefont {L.}~\bibnamefont
  {Lindblom}},\ }\href {\doibase 10.1103/PhysRevD.97.123019} {\bibfield
  {journal} {\bibinfo  {journal} {Phys. Rev. D}\ }\textbf {\bibinfo {volume}
  {97}},\ \bibinfo {pages} {123019} (\bibinfo {year} {2018})},\ \Eprint
  {http://arxiv.org/abs/1804.04072} {arXiv:1804.04072 [astro-ph.HE]}
  \BibitemShut {NoStop}%
\bibitem [{\citenamefont {Annala}\ \emph
  {et~al.}(2020{\natexlab{b}})\citenamefont {Annala}, \citenamefont {Gorda},
  \citenamefont {Kurkela}, \citenamefont {N\"attil\"a},\ and\ \citenamefont
  {Vuorinen}}]{Annala:2019puf}%
  \BibitemOpen
  \bibfield  {author} {\bibinfo {author} {\bibfnamefont {E.}~\bibnamefont
  {Annala}}, \bibinfo {author} {\bibfnamefont {T.}~\bibnamefont {Gorda}},
  \bibinfo {author} {\bibfnamefont {A.}~\bibnamefont {Kurkela}}, \bibinfo
  {author} {\bibfnamefont {J.}~\bibnamefont {N\"attil\"a}}, \ and\ \bibinfo
  {author} {\bibfnamefont {A.}~\bibnamefont {Vuorinen}},\ }\href {\doibase
  10.1038/s41567-020-0914-9} {\bibfield  {journal} {\bibinfo  {journal} {Nature
  Phys.}\ }\textbf {\bibinfo {volume} {16}},\ \bibinfo {pages} {907} (\bibinfo
  {year} {2020}{\natexlab{b}})},\ \Eprint {http://arxiv.org/abs/1903.09121}
  {arXiv:1903.09121 [astro-ph.HE]} \BibitemShut {NoStop}%
\bibitem [{\citenamefont {Kumar}\ \emph {et~al.}(2023)\citenamefont {Kumar},
  \citenamefont {Mishra},\ and\ \citenamefont {Malik}}]{Kumar:2021hzo}%
  \BibitemOpen
  \bibfield  {author} {\bibinfo {author} {\bibfnamefont {D.}~\bibnamefont
  {Kumar}}, \bibinfo {author} {\bibfnamefont {H.}~\bibnamefont {Mishra}}, \
  and\ \bibinfo {author} {\bibfnamefont {T.}~\bibnamefont {Malik}},\ }\href
  {\doibase 10.1088/1475-7516/2023/02/015} {\bibfield  {journal} {\bibinfo
  {journal} {JCAP}\ }\textbf {\bibinfo {volume} {02}},\ \bibinfo {pages} {015}
  (\bibinfo {year} {2023})},\ \Eprint {http://arxiv.org/abs/2110.00324}
  {arXiv:2110.00324 [hep-ph]} \BibitemShut {NoStop}%
\bibitem [{\citenamefont {Miller}\ \emph {et~al.}(2021)\citenamefont {Miller}
  \emph {et~al.}}]{Miller:2021qha}%
  \BibitemOpen
  \bibfield  {author} {\bibinfo {author} {\bibfnamefont {M.~C.}\ \bibnamefont
  {Miller}} \emph {et~al.},\ }\href {\doibase 10.3847/2041-8213/ac089b}
  {\bibfield  {journal} {\bibinfo  {journal} {Astrophys. J. Lett.}\ }\textbf
  {\bibinfo {volume} {918}},\ \bibinfo {pages} {L28} (\bibinfo {year}
  {2021})},\ \Eprint {http://arxiv.org/abs/2105.06979} {arXiv:2105.06979
  [astro-ph.HE]} \BibitemShut {NoStop}%
\bibitem [{\citenamefont {Abbott}\ \emph {et~al.}(2019)\citenamefont {Abbott}
  \emph {et~al.}}]{LIGOScientific:2018hze}%
  \BibitemOpen
  \bibfield  {author} {\bibinfo {author} {\bibfnamefont {B.~P.}\ \bibnamefont
  {Abbott}} \emph {et~al.} (\bibinfo {collaboration} {LIGO Scientific,
  Virgo}),\ }\href {\doibase 10.1103/PhysRevX.9.011001} {\bibfield  {journal}
  {\bibinfo  {journal} {Phys. Rev. X}\ }\textbf {\bibinfo {volume} {9}},\
  \bibinfo {pages} {011001} (\bibinfo {year} {2019})},\ \Eprint
  {http://arxiv.org/abs/1805.11579} {arXiv:1805.11579 [gr-qc]} \BibitemShut
  {NoStop}%
\bibitem [{\citenamefont {Riley}\ \emph {et~al.}(2019)\citenamefont {Riley}
  \emph {et~al.}}]{Riley:2019yda}%
  \BibitemOpen
  \bibfield  {author} {\bibinfo {author} {\bibfnamefont {T.~E.}\ \bibnamefont
  {Riley}} \emph {et~al.},\ }\href {\doibase 10.3847/2041-8213/ab481c}
  {\bibfield  {journal} {\bibinfo  {journal} {Astrophys. J. Lett.}\ }\textbf
  {\bibinfo {volume} {887}},\ \bibinfo {pages} {L21} (\bibinfo {year}
  {2019})},\ \Eprint {http://arxiv.org/abs/1912.05702} {arXiv:1912.05702
  [astro-ph.HE]} \BibitemShut {NoStop}%
\bibitem [{\citenamefont {Miller}\ \emph {et~al.}(2019)\citenamefont {Miller}
  \emph {et~al.}}]{Miller:2019cac}%
  \BibitemOpen
  \bibfield  {author} {\bibinfo {author} {\bibfnamefont {M.~C.}\ \bibnamefont
  {Miller}} \emph {et~al.},\ }\href {\doibase 10.3847/2041-8213/ab50c5}
  {\bibfield  {journal} {\bibinfo  {journal} {Astrophys. J. Lett.}\ }\textbf
  {\bibinfo {volume} {887}},\ \bibinfo {pages} {L24} (\bibinfo {year}
  {2019})},\ \Eprint {http://arxiv.org/abs/1912.05705} {arXiv:1912.05705
  [astro-ph.HE]} \BibitemShut {NoStop}%
\bibitem [{\citenamefont {Riley}\ \emph {et~al.}(2021)\citenamefont {Riley}
  \emph {et~al.}}]{Riley:2021pdl}%
  \BibitemOpen
  \bibfield  {author} {\bibinfo {author} {\bibfnamefont {T.~E.}\ \bibnamefont
  {Riley}} \emph {et~al.},\ }\href {\doibase 10.3847/2041-8213/ac0a81}
  {\bibfield  {journal} {\bibinfo  {journal} {Astrophys. J. Lett.}\ }\textbf
  {\bibinfo {volume} {918}},\ \bibinfo {pages} {L27} (\bibinfo {year}
  {2021})},\ \Eprint {http://arxiv.org/abs/2105.06980} {arXiv:2105.06980
  [astro-ph.HE]} \BibitemShut {NoStop}%
\bibitem [{\citenamefont {Doneva}\ \emph {et~al.}(2013)\citenamefont {Doneva},
  \citenamefont {Gaertig}, \citenamefont {Kokkotas},\ and\ \citenamefont
  {Kr\"uger}}]{Doneva:2013zqa}%
  \BibitemOpen
  \bibfield  {author} {\bibinfo {author} {\bibfnamefont {D.~D.}\ \bibnamefont
  {Doneva}}, \bibinfo {author} {\bibfnamefont {E.}~\bibnamefont {Gaertig}},
  \bibinfo {author} {\bibfnamefont {K.~D.}\ \bibnamefont {Kokkotas}}, \ and\
  \bibinfo {author} {\bibfnamefont {C.}~\bibnamefont {Kr\"uger}},\ }\href
  {\doibase 10.1103/PhysRevD.88.044052} {\bibfield  {journal} {\bibinfo
  {journal} {Phys. Rev. D}\ }\textbf {\bibinfo {volume} {88}},\ \bibinfo
  {pages} {044052} (\bibinfo {year} {2013})},\ \Eprint
  {http://arxiv.org/abs/1305.7197} {arXiv:1305.7197 [astro-ph.SR]} \BibitemShut
  {NoStop}%
\bibitem [{\citenamefont {Pradhan}\ \emph {et~al.}(2022)\citenamefont
  {Pradhan}, \citenamefont {Chatterjee}, \citenamefont {Lanoye},\ and\
  \citenamefont {Jaikumar}}]{Pradhan:2022vdf}%
  \BibitemOpen
  \bibfield  {author} {\bibinfo {author} {\bibfnamefont {B.~K.}\ \bibnamefont
  {Pradhan}}, \bibinfo {author} {\bibfnamefont {D.}~\bibnamefont {Chatterjee}},
  \bibinfo {author} {\bibfnamefont {M.}~\bibnamefont {Lanoye}}, \ and\ \bibinfo
  {author} {\bibfnamefont {P.}~\bibnamefont {Jaikumar}},\ }\href {\doibase
  10.1103/PhysRevC.106.015805} {\bibfield  {journal} {\bibinfo  {journal}
  {Phys. Rev. C}\ }\textbf {\bibinfo {volume} {106}},\ \bibinfo {pages}
  {015805} (\bibinfo {year} {2022})},\ \Eprint
  {http://arxiv.org/abs/2203.03141} {arXiv:2203.03141 [astro-ph.HE]}
  \BibitemShut {NoStop}%
\bibitem [{\citenamefont {Yoshida}\ and\ \citenamefont
  {Kojima}(1997)}]{Yoshida:1997bf}%
  \BibitemOpen
  \bibfield  {author} {\bibinfo {author} {\bibfnamefont {S.}~\bibnamefont
  {Yoshida}}\ and\ \bibinfo {author} {\bibfnamefont {Y.}~\bibnamefont
  {Kojima}},\ }\href {\doibase 10.1093/mnras/289.1.117} {\bibfield  {journal}
  {\bibinfo  {journal} {Mon. Not. Roy. Astron. Soc.}\ }\textbf {\bibinfo
  {volume} {289}},\ \bibinfo {pages} {117} (\bibinfo {year} {1997})},\ \Eprint
  {http://arxiv.org/abs/gr-qc/9705081} {arXiv:gr-qc/9705081} \BibitemShut
  {NoStop}%
\bibitem [{\citenamefont {Kokkotas}\ \emph {et~al.}(2001)\citenamefont
  {Kokkotas}, \citenamefont {Apostolatos},\ and\ \citenamefont
  {Andersson}}]{Kokkotas:1999mn}%
  \BibitemOpen
  \bibfield  {author} {\bibinfo {author} {\bibfnamefont {K.~D.}\ \bibnamefont
  {Kokkotas}}, \bibinfo {author} {\bibfnamefont {T.~A.}\ \bibnamefont
  {Apostolatos}}, \ and\ \bibinfo {author} {\bibfnamefont {N.}~\bibnamefont
  {Andersson}},\ }\href {\doibase 10.1046/j.1365-8711.2001.03945.x} {\bibfield
  {journal} {\bibinfo  {journal} {Mon. Not. Roy. Astron. Soc.}\ }\textbf
  {\bibinfo {volume} {320}},\ \bibinfo {pages} {307} (\bibinfo {year}
  {2001})},\ \Eprint {http://arxiv.org/abs/gr-qc/9901072} {arXiv:gr-qc/9901072}
  \BibitemShut {NoStop}%
\bibitem [{\citenamefont {Pradhan}\ and\ \citenamefont
  {Chatterjee}(2021)}]{Pradhan:2020amo}%
  \BibitemOpen
  \bibfield  {author} {\bibinfo {author} {\bibfnamefont {B.~K.}\ \bibnamefont
  {Pradhan}}\ and\ \bibinfo {author} {\bibfnamefont {D.}~\bibnamefont
  {Chatterjee}},\ }\href {\doibase 10.1103/PhysRevC.103.035810} {\bibfield
  {journal} {\bibinfo  {journal} {Phys. Rev. C}\ }\textbf {\bibinfo {volume}
  {103}},\ \bibinfo {pages} {035810} (\bibinfo {year} {2021})},\ \Eprint
  {http://arxiv.org/abs/2011.02204} {arXiv:2011.02204 [astro-ph.HE]}
  \BibitemShut {NoStop}%
\end{thebibliography}%

\end{document}